\documentclass[usenatbib]{mnras}
\usepackage{hyperref}
\usepackage{url}
\usepackage{graphicx}
\usepackage{amssymb}
\usepackage{amsmath}
\usepackage[T1]{fontenc}
\usepackage{ae,aecompl}

\usepackage{etoolbox}
\makeatletter
\newcount\c@additionalboxlevel
\newcount\c@maxboxlevel
\patchcmd\@combinedblfloats{\box\@outputbox}{%
  \stepcounter{additionalboxlevel}%
  \box\@outputbox
}{}{\errmessage{\noexpand\@combinedblfloats could not be patched}}

\AtBeginShipout{%
  \ifnum\value{additionalboxlevel}>\value{maxboxlevel}%
    \typeout{Warning: maxboxlevel might be too small, increase to %
      \the\value{additionalboxlevel}%
    }%
  \fi 
  \@whilenum\value{additionalboxlevel}<\value{maxboxlevel}\do{%
    \typeout{* Additional boxing of page `\thepage'}%
    \setbox\AtBeginShipoutBox=\hbox{\copy\AtBeginShipoutBox}%
    \stepcounter{additionalboxlevel}%
  }%
  \setcounter{additionalboxlevel}{0}%
}
\makeatother

\newcommand{\unit}[2]{\ensuremath{\textrm{#1}^{#2}}}

\title[New views of the distant stellar halo]{New views of the distant stellar halo}
\author[R.E. Sanderson et al.]{Robyn E. Sanderson,$^{1,2}$\thanks{robyn@caltech.edu}\thanks{NSF Astronomy and Astrophysics Postdoctoral Fellow} 
Amy Secunda,$^{1}$ 
Kathryn V. Johnston,$^{1}$
 John J. Bochanski$^{3}$ \\
$^{1}$Department of Astronomy, Columbia University, 550 W 120th St, New York, NY 10027\\
$^{2}$TAPIR, Caltech, MC 301-17, 1200 E. California Blvd., Pasadena, CA 91125\\
$^{3}$Department of Chemistry, Biochemistry and Physics, Rider University, 2083 Lawrenceville Road, Lawrenceville, NJ 08648}

\date{Accepted XXX. Received YYY; in original form ZZZ}

\pubyear{2016}

\begin{document}
\label{firstpage}
\pagerange{\pageref{firstpage}--\pageref{lastpage}}
\maketitle

\begin{abstract}
Currently only a small number of Milky Way (MW) stars are known to exist beyond 100 kpc from the Galactic center.  Though the distribution of these stars in the outer halo is believed to be sparse, they can provide evidence of more recent accretion events than in the inner halo and help map out the MW's dark matter halo to its virial radius. We have re-examined the outermost regions of 11 existing stellar halo models with two synthetic surveys: one mimicking present-day searches for distant M giants and another mimicking RR Lyrae (RRLe) projections for LSST. Our models suggest that color and proper motion cuts currently used to select M giant candidates for follow-up successfully remove nearly all halo dwarf self-contamination and are useful for focusing observations on distant M giants, of which there are thousands to tens of thousands beyond 100 kpc in our models. We likewise expect that LSST will identify comparable numbers of RRLe at these distances. We demonstrate that several observable properties of both tracers, such as proximity of neighboring stars, proper motions, and distances (for RRLe) could help us separate different accreted structures from one another. We also discuss prospects for using ratios of M giants to RRLe as a proxy for accretion time, which in the future could provide new constraints on the recent accretion history of our Galaxy.
\end{abstract}


\section{Introduction}  
\label{sec:intro}

Stellar halos of spiral galaxies typically contain of order a few percent of the total number of stars associated with their host dark matter halos, spread over spatial scales ten times larger than the disks that they surround. Hence they are insignificant in terms of understanding the bulk of baryonic material that has occurred throughout the history of the Universe, and their extremely low densities (and corresponding surface brightness) makes them in any case difficult to study. However, two properties of stellar halos make them uniquely interesting. First, it is here that it is most productive to search for stars that were {\it not} formed in the current host halo, but rather accreted from other objects. Hence, the properties of the stellar populations of the halo can tell us something both about the accretion histories of galaxies, as well as the properties of the (now-dead) dwarf galaxies that formed them. Secondly, the low total mass and vast spatial scales that halos stars explore makes them powerful probes of the mass and structure of dark matter halos that surround all galaxies.

The production of vast catalogues of faint stars around our own \citep[see][for a review]{ivezic12} and other \citep{ferguson02} galaxies have for the first time allowed the global structure of several stellar halos to be convincingly mapped \citep{ibata14}. These studies have also revealed the presence of a significant contribution of  substructure in space \citep{ferguson02,newberg02,belokurov06} and velocity \citep{schlaufman09,gilbert09} which can be attributed to the hierarchical nature of galaxy formation \citep{bullock01} —-- the substructures are the debris from the destruction of infalling dwarf galaxies. Comparisons with concurrent theoretical work suggests broad consistency of these observations with the expectations for the scales, structure, and frequency of substructure in stellar halos built within the $\Lambda$CDM paradigm \citep{bell08,bell10,xue11}. Some of these substructures have been exploited as probes of the underlying gravitational potential  \citep{2004Helmi,2005Johnston,2005Law,2009Willett,2009Law,2010Koposov,2010Newberg,2010Law,2012MNRAS.424L..16L,2013ApJ...776...26S,2013Vera-Ciro,kuepper15,pearson15}.

A remaining frontier in this field is the mapping and interpretation of the outermost regions of galactic halos all the way out to the virial radius ($\sim$300kpc for a Milky-Way-mass galaxy). M31 is the only galaxy in the Universe for which a global map has been made on these scales, reaching to $\sim150$kpc \citep{ibata14}. In contrast, for the Milky Way, the views of the stellar halo afforded by Main Sequence Turnoff Stars selected from SDSS extend to $\sim$40kpc and the M-giants extracted from 2MASS reach distances of less than 100kpc \citep{belokurov06,majewski03}. In the next few years, data releases from the Gaia satellite promise to fill in these maps with vast numbers of stars and additional dimensions of information, but Gaia's magnitude limit of roughly $V\sim 20$ again restricts sensitivity to within roughly 100 kpc of the Galactic center for bright giant tracers.

The number of Milky Way stars known to lie beyond 100 kpc from the Galactic center is still very small, but steadily growing. Large areal surveys with deep, precise photometry have been critically important to identifying relatively rare, but luminous, halo stars.  Two classes of stars are bright enough to be observed beyond 120 kpc with current surveys: blue horizontal branch (BHB) stars and M giants.  \cite{2012MNRAS.425.2840D} selected a sample of seven spectroscopically confirmed BHB stars in SDSS with distances of 80 kpc $< d <$ 150 kpc.  At distances greater than 150 kpc, only M giants are bright enough to be readily observed in modern day surveys.  \cite{2014AJ....147...76B} assembled a sample of nearly 500 M giant candidate stars with optical and infrared photometry from SDSS and UKIDSS.  The M giants in the \citeauthor{2014AJ....147...76B} sample can be seen from 30 to $\sim$ 300 kpc, making them the first to probe the stellar content of the Milky Way near the virial radius.  Unfortunately, photometry and the lack of proper motions are not sufficient to identify M giants, making spectroscopy necessary.  Despite an estimated contamination rate near 80\%, \cite{2014ApJ...790L...5B} have already spectroscopically confirmed two distant M giants, with estimated distances over 200 kpc.  The most distant M giant known, ULAS J001535.72+015549.6 has a distance of 274 $\pm$ 74 kpc.  This sample has yielded 10 confirmed M giants to date, most being part of the Sagittarius dwarf galaxy remnant. 

Further into the future, we can anticipate dramatic additions to these outer halo detections as the Large Synoptic Survey Telescope \citep{ivezic08} produces catalogs of stars as faint as $g=24.5$ in a single pointing, corresponding to a distance limit of $\sim$ 50 kpc for a main-sequence turnoff (MSTO) star and $\sim$ 600 kpc for an RR Lyrae (RRL) star. After 5 years the co-added data will reach $g=27$, out to $\sim$ 300 kpc for an MSTO and $\sim$ 3 Mpc for an RRL. It is as yet unclear what to expect in this regime. Model stellar halos show them becoming more dominated by substructure at larger galactocentric radii as the dynamical timescales become comparable to the age of the Universe and the debris from the few recent accretion events has little time to phase-mix away  \citep{johnston08}. M31’s stellar halo extends at least to 150 kpc and is richly substructured \citep{ibata14}, but the stochastic nature of hierarchical structure formation ensures a vast variety in stellar halo structures, especially on these spatial scales, so the Milky Way's stellar halo could differ dramatically. 

	Even if the populations of stars in the outermost halo prove to be very sparse, they will have some important implications: they will provide a view of accretion (or perhaps lack of accretion?) in a new and unique regime, likely more sensitive recent events; and they will provide dynamical tracers to map the dark matter halo all the way out to the virial radius. This paper is motivated by the steadily growing number of stars known to be beyond 100kpc from the Galactic Center, as well as the longer-term prospects for LSST, to re-examine the outermost reaches of 11 existing stellar halo models \citep{2005ApJ...635..931B} in order to explore our expectations for these populations in a little more detail.  In particular, this study looks at two different types of stars that might be selected in current and future stellar catalogues: color-selected M giants and time-domain selected RR Lyrae. For each tracer, we examine the trends and diversity in numbers of stars and properties of objects from which they came in the models. We also discuss the likelihood of  being able to make associations between stars from their observed properties - associations that will increase our ability to reconstruct both the full accretion history of our Galaxy as well as the structure of its dark matter halo.

This paper is organized as follows: in Section 2 we describe the computational tools (the mock halos and {\sc Galaxia}) used to generate synthetic stellar populations in the outer halo; in Section 3 we use these synthetic surveys to discuss expectations for present-day searches for distant M giants and in Section 4 we discuss future prospects for RR Lyrae. In Section 5 we discuss the possibility of using ratios of these two tracers to reconstruct the Milky Way's accretion history. In Section 6 we summarize our findings, draw some conclusions, and indicate some directions for future work.

\section{Toolbox}
\label{sec:tools}
We use the set of publicly available\footnote{\url{http://user.astro.columbia.edu/~kvj/halos/}} mock stellar halos from \citet{2005ApJ...635..931B}, and look at M giant and RR Lyrae stellar tracers by generating synthetic surveys with {\sc Galaxia}\footnote{\url{http://galaxia.sourceforge.net}} \citep{2011ApJ...730....3S}. Table 1 summarizes the color and magnitude cuts used for each of these tracers.

\subsection{Mock stellar halo simulations}
\label{subsec:tools:bjhalos}
The stellar halo models described in \citet{2005ApJ...635..931B} were built entirely from accretion events drawn from histories representing random realizations of the formation of a Milky-Way-type galaxy in a $\Lambda$CDM Universe.
The phase-space structure in the models was constructed by superposing the final positions and velocities of particles at the end point of individual N-body simulations of dwarf galaxies disrupting around a parent galaxy matching the cosmological model accretion event history. All subhalos crossing within the virial radius of the parent galaxy (282 kpc at present day) were tracked. The 100,000 massive particles in each of these simulated objects had equal dark matter masses and were given (varying) associated mass-to-light ratios in such a way that the luminous material reproduced the structural scaling relations observed for Local Group dwarfs. 
The level of resolution of the phase space structure of resulting stellar halo was increased by introducing an additional 100,000 mass-less test particles with the same energy distribution as the 20\% most-tightly bound dark matter particles.  
Star formation histories were assigned to the star-particles within each accreted dwarf using a simple leaky-accreting-box model of star formation and chemical enrichment that was abruptly truncated upon accretion \citep{robertson05,font06}.

Two key attributes of the \citet{2005ApJ...635..931B} models should be borne in mind throughout the rest of this paper.
First, the models only represent the portion of halo stars that were formed in and subsequently {\it accreted} from other dark matter halos. 
Fully self-consistent, cosmological hydrodynamical simulations of the formation of Milky Way-type galaxies have typically been found to contain an additional population beyond their galactic disks that formed within the main galaxy's own dark matter halo \citep{abadi06,zolotov09,font11,tissera13,pillepich15}
However, while these models typically differ on the percentage and radial distribution of stellar halo stars formed this way, none predict a significant portion beyond $\sim$50kpc from the Galactic center. 
Hence we anticipate our consideration only of accreted populations to be a valid simplification for the study of the outer halo.

Second, the models were built to match the number and luminosities of the dozen contemporaneously-known satellites of the Milky Way, and hence do not contain the more numerous population of less luminous ultra-faint dwarf (UFD) satellites that have been discovered since that time \citep[e.g.][]{willman05,belokurov06b,drlicawagner15}, nor stars from the ancestral analogues of the surviving UFD's that would have been disrupted at earlier times.
However, we would expect these many smaller galaxies to contribute stars that might occupy unique corners of stellar populations and abundance space. In particular, these galaxies would be too metal-poor to have any M giants associated with them, so their absence in the models does not affect our comparison with the UKIDSS survey.
On the other hand, their contributions could alter the size and phase-space structure of the RR Lyrae halo.

Given these limitations, our examination of these models is intended as indicative, rather than  predictive.
This would be true for any stellar halo models that could be examined.
While there is very broad agreement among the different models (and across techniques) that stars in the halo beyond 50kpc are most likely accreted, there is no clear consensus on the amount or type of material or expected phase-space structure \citep{2005ApJ...635..931B,delucia08,cooper10,helmi11}.
Part of this apparent disagreement is undoubtedly due to the stochastic nature of hierarchical structure formation (as reflected in this paper in Figures \ref{fig:Mgiants_allsky_present} and \ref{fig:rrl_asp}). However, some can be attributed to differences between representations of either the physics or objects in the Universe in the models themselves.\citep[For one example, see discussion in][on how the presence or absence of a disk potential can change stellar halo properties in models.]{bailin14}
These differences have been driven by computational rather than physical considerations---models with fully self-consistent representations of the range of physical processes needed to be fully predictive and with high enough resolution to compare to star-count studies do not yet exist in large numbers, although they are on the horizon \citep{2016ApJ...827L..23W}.

\begin{table*} 
\begin{center}
\caption{Stellar tracers used in this work.}
\begin{tabular}{p{0.5in}ccp{1.25in}p{1.0in}p{1.6in}}
\hline
Population & Abs. mag & App. Mag range & Color range & Approx. distance range (kpc) & Note \\
\hline
RR Lyr & $-1.0<M_{K_s}<0.1$ & $10<m_{K_s}<24.5$ & $0.13<J-K_s<0.363$ & $d<600$ & as in \citet{2011ApJ...728..106S}\\
M giants & $0.05<M_J<0.25$  & $12.5<J<18.5$ & $0.05<H-K<0.4$ & $30<d<300 $ & as in \citet{2014ApJ...790L...5B}\\
\hline
\end{tabular}
\label{tbl:tracers}
\end{center}
\end{table*}

\subsection{Synthetic surveys with {\sc Galaxia}}
\label{subsec:tools:galaxia}
To generate synthetic surveys of the mock stellar halos, we use the public software package {\sc Galaxia} \citep{2011ApJ...730....3S}. This code resamples the particle representations of the building blocks of a given mock halo to generate star particles, following the luminosity weights assigned to each N-body particle as discussed in Section \ref{subsec:tools:bjhalos} and presuming stellar populations represented by the Padova isochrones \citep{1994A&AS..106..275B,2008A&A...482..883M}. The code returns a catalog of ``stars'' that fall inside a color-magnitude box and sky area set by the user. We refer the interested reader to \citet{2011ApJ...730....3S} for the details of the resampling process, but note that each individual building block in a given halo is resampled individually when the mock catalog is generated. The code performs the phase-space resampling in two 3D subspaces (positions and velocities) rather than in the full 6D space, and considering each building block individually ensures that the sampler does not overestimate their configuration-space thickness or velocity dispersion through confusion between streams from multiple objects.

\begin{figure*}
\begin{tabular}{cc}
\includegraphics[width=0.45\textwidth]{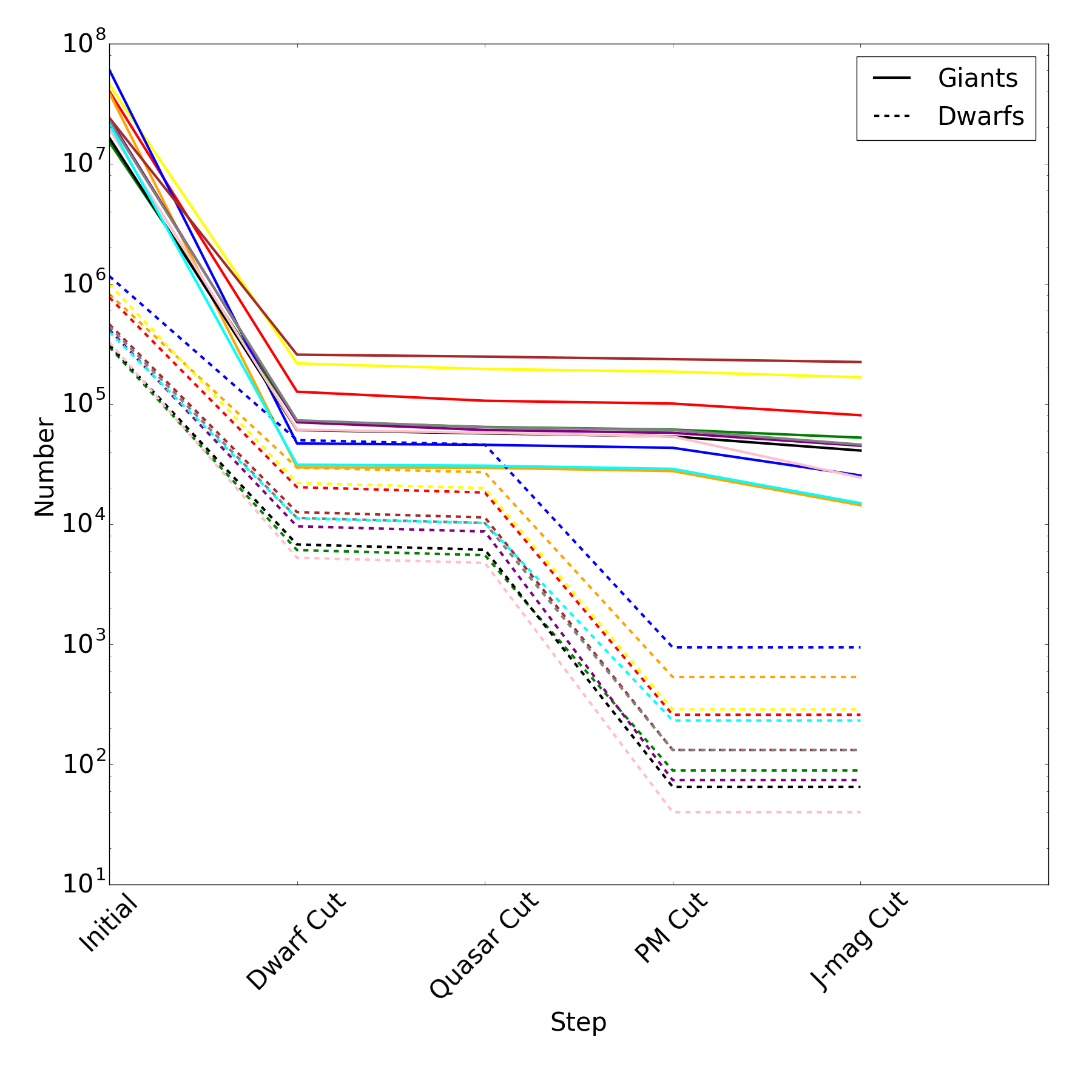} \qquad \includegraphics[width=0.45\textwidth]{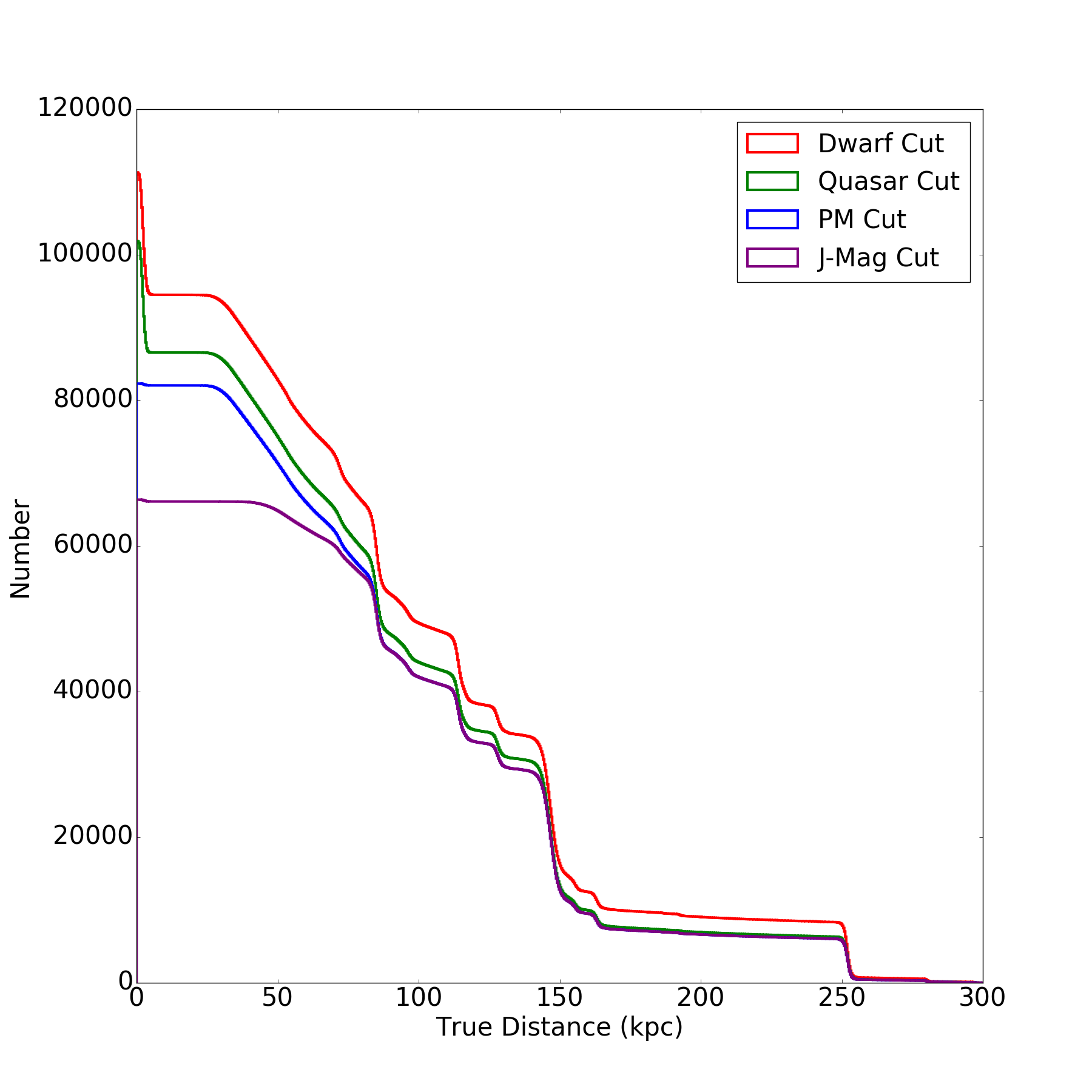}
\end{tabular}
\caption{Left: The number of M Giants (represented by the solid lines) and dwarfs (represented by the dashed lines) remaining in the error-convolved sample after each cut is made. Each color represents a different mock halo. Right: The cumulative, from the outside in, number of stars remaining in the error-convolved sample as a function of true distance after each additional cut is made. In this panel the different colors represent the successive selections on the sample, made in the order outlined in Section \ref{subsec:mgiants-present:select}.}
\label{fig:prog_cuts}
\end{figure*}

\section{Distant M giants: Mocking a present-day survey}
\label{sec:mgiants-present}
Searches are ongoing for distant M giants in wide-field surveys, and have already yielded the two most distant known stars in the Milky Way \citep{2014ApJ...790L...5B}. Where do these stars come from? Are they more likely to be in bound structures or in tidal streams? What distance distribution do we expect? Where is the edge of the stellar halo when looking at this tracer?

To explore these questions we generated synthetic surveys of a box in color-magnitude space containing M giants, convolved these synthetic stars with photometric and proper motion errors corresponding to present and future datasets, and selected the distant M giants from the mock catalogue with the same infrared and visible color cuts as the candidate sample in \citet{2014ApJ...790L...5B}, for each of the 11 different mock stellar halos in \citet{2005ApJ...635..931B}.

\begin{figure*}
\begin{tabular}{cc}
\includegraphics[width=0.45\textwidth]{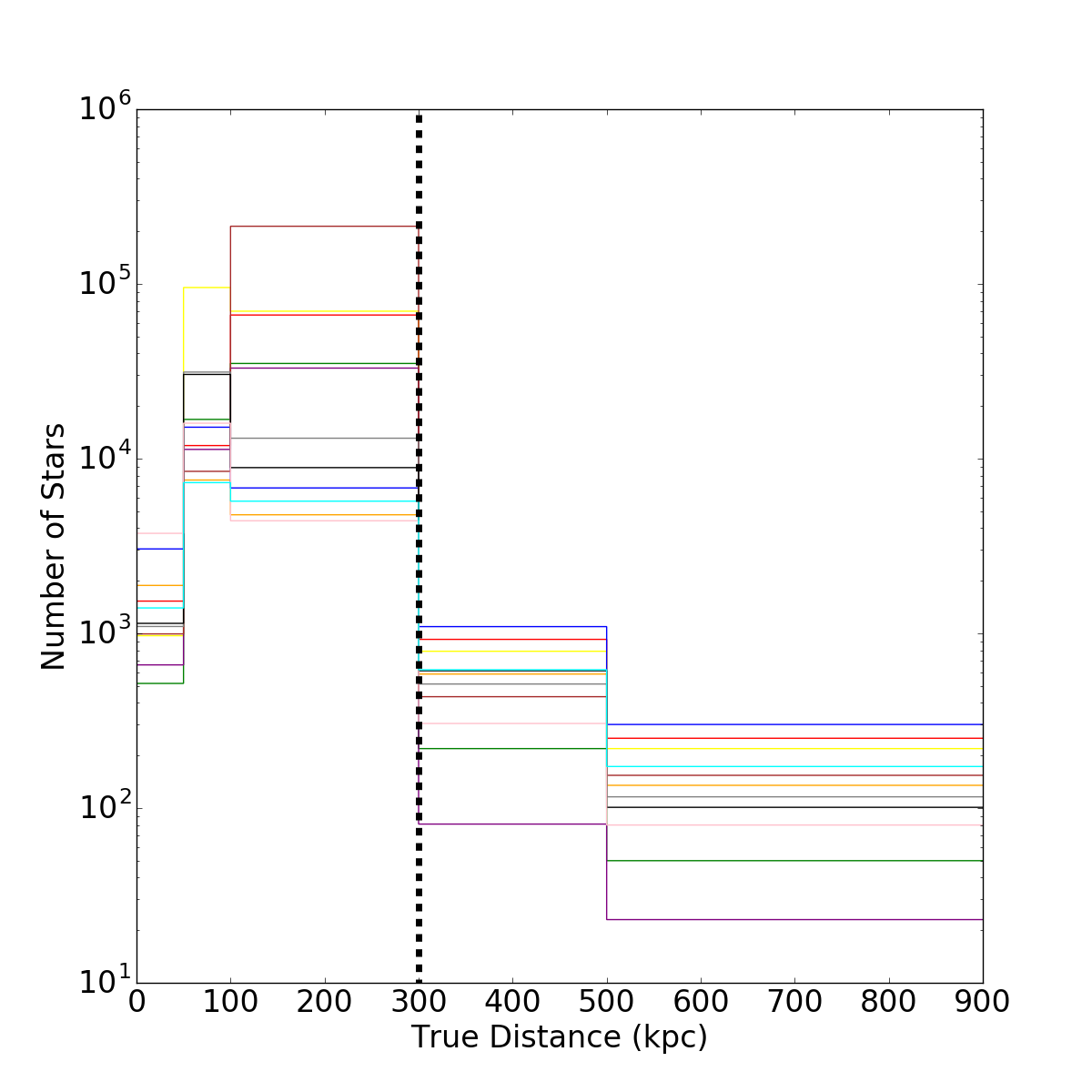} & \includegraphics[width=0.45\textwidth]{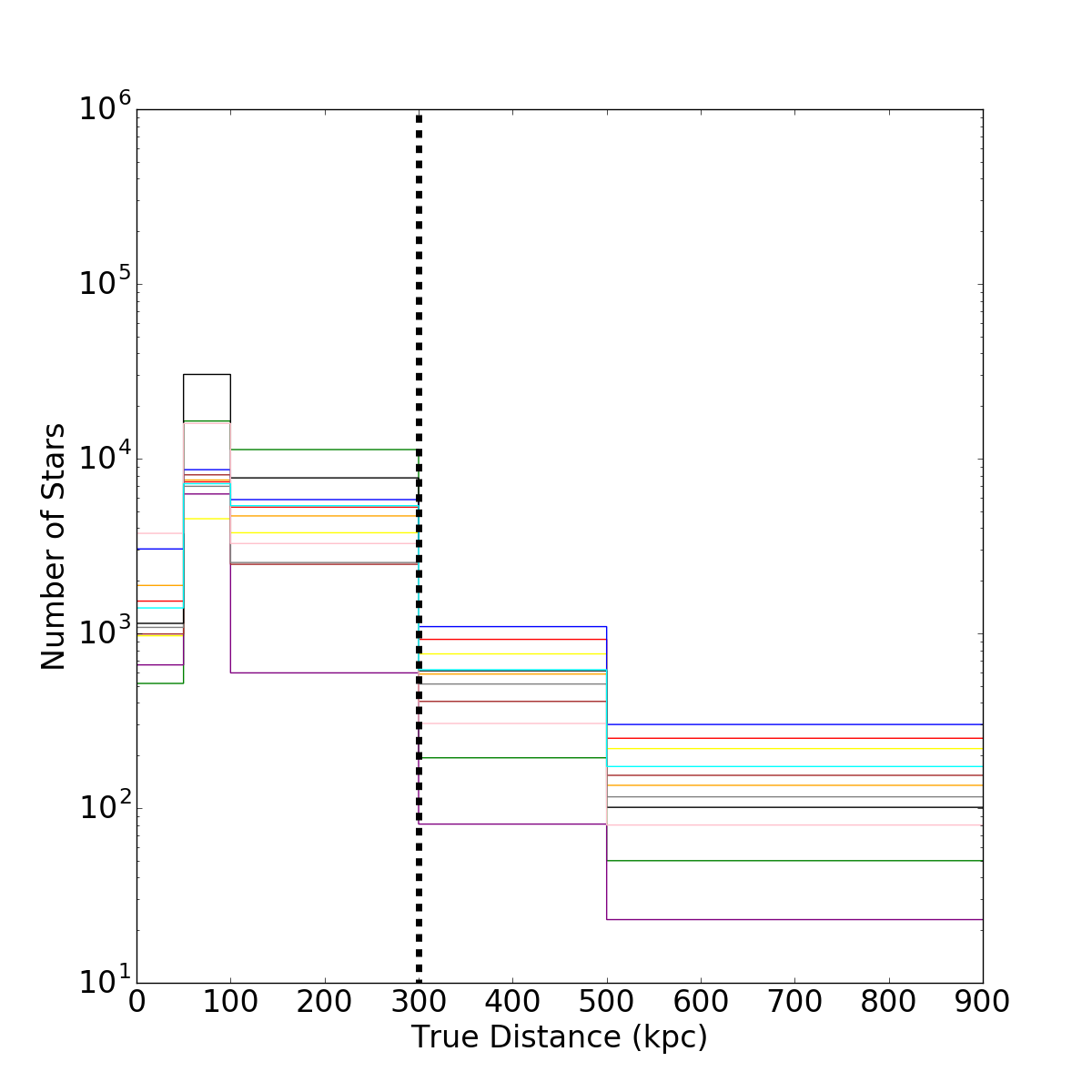}
\end{tabular}
\caption{Number of selected M giants for each of the mock halos as a function of true distance for all substructures (left) and excluding bound satellites (right). Each color represents a different mock halo. The dashed vertical line represents the virial radius.}
\label{fig:mg_cumul_present}
\end{figure*}

\begin{figure*}
\begin{tabular}{cc}
\includegraphics[width=0.5\textwidth]{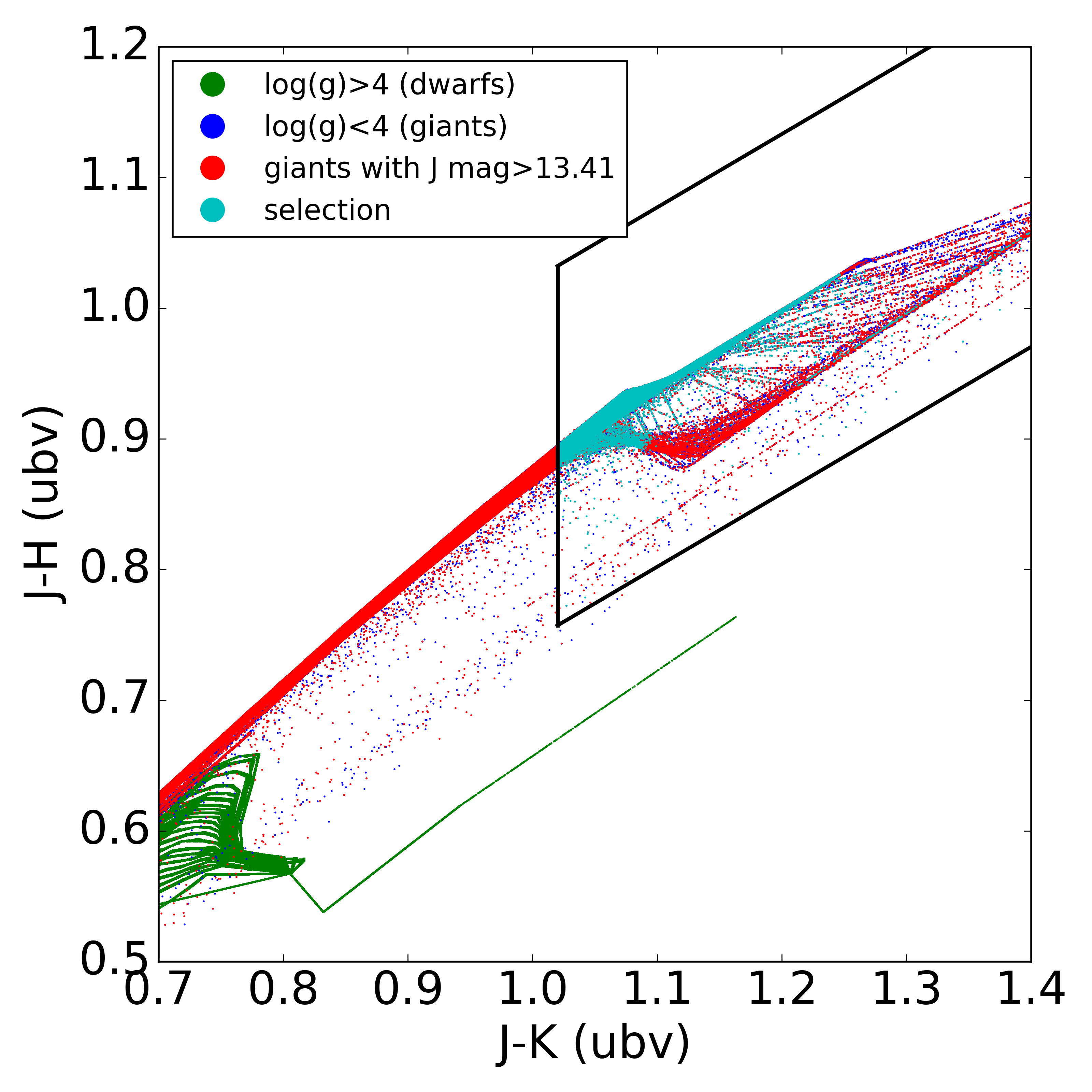} & \includegraphics[width=0.5\textwidth]{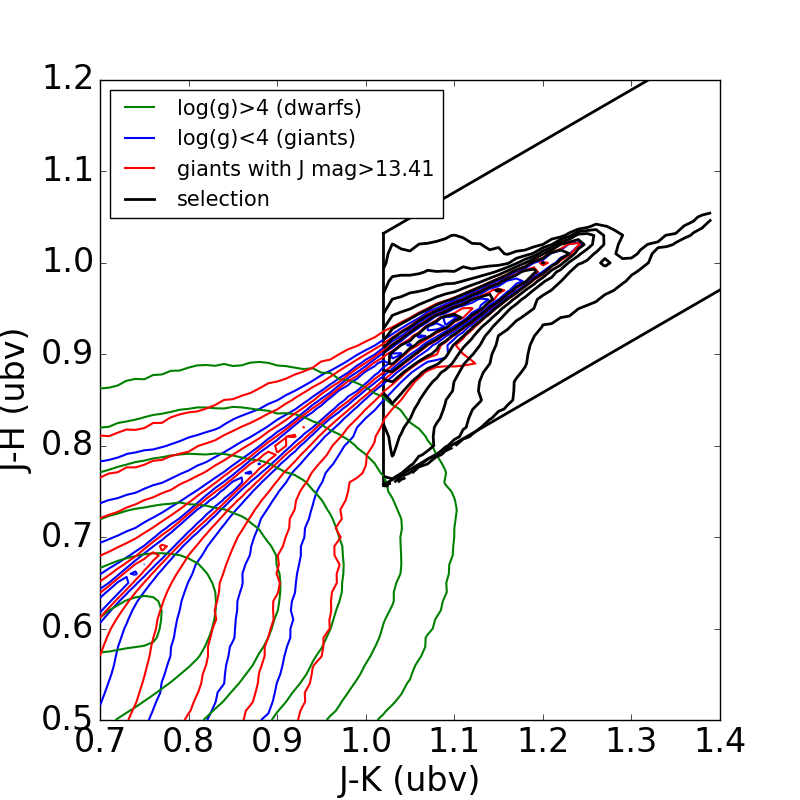}
\end{tabular}
\caption{Example of selection of giant stars in infrared color-color space for all 11 halos stacked together (see Section \ref{subsec:mgiants-present:select}). In both panels stars inside the black box are selected. Green points/contours show synthetic stars with $\log g>4$ (dwarfs), blue points/contours are synthetic stars with $\log g<4$ (giants), red indicates giants with J magnitudes greater than 13.41, and cyan points/black contours show stars selected based on the criteria outlined in Section \ref{subsec:mgiants-present:select}. Left: the selection operating on the synthetic survey without error convolution. Right: the same selection applied to the error-convolved synthetic survey.}
\label{fig:Mgiants_cc}
\end{figure*}

\begin{figure*}
\begin{tabular}{cc}
\includegraphics[width=0.5\textwidth]{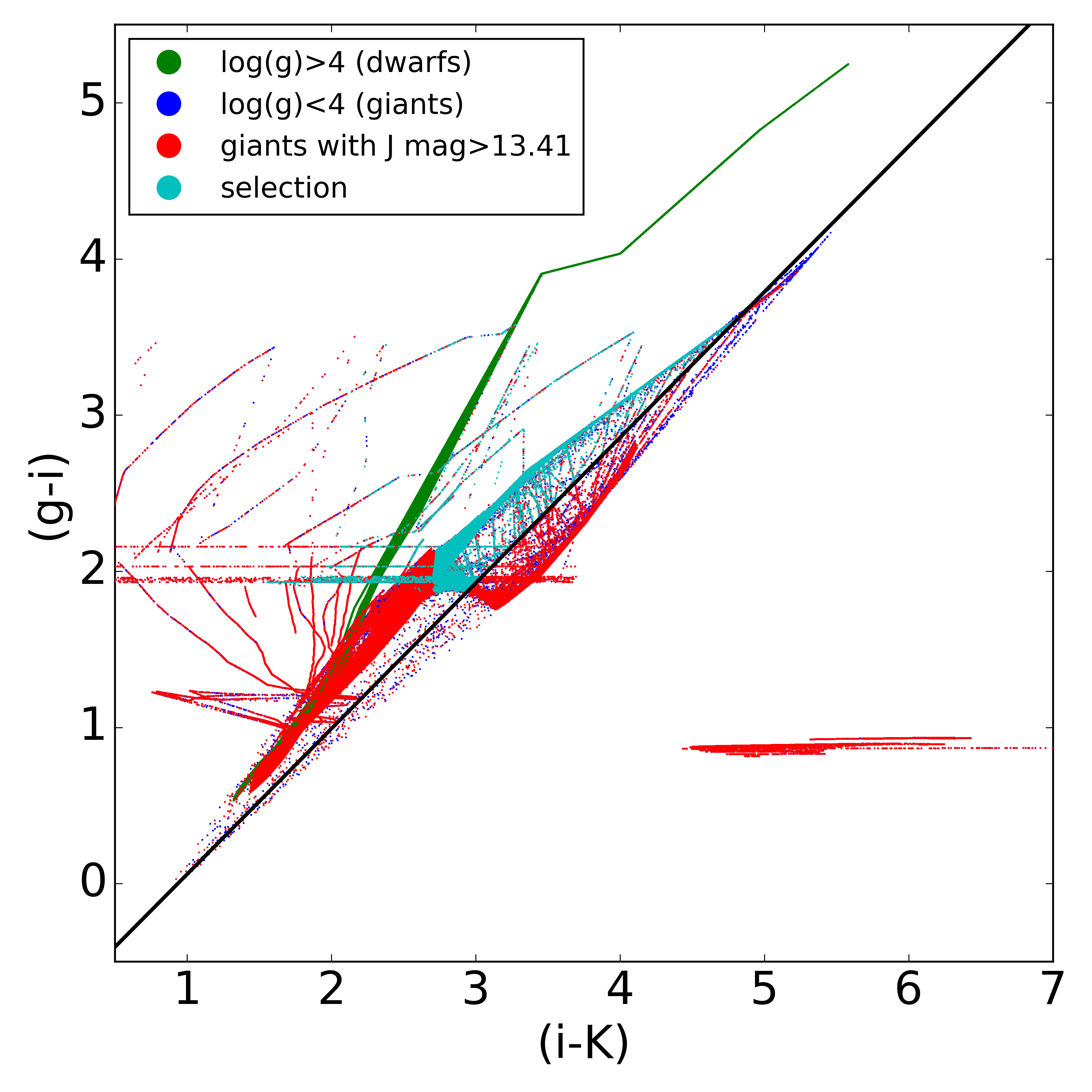} & \includegraphics[width=0.5\textwidth]{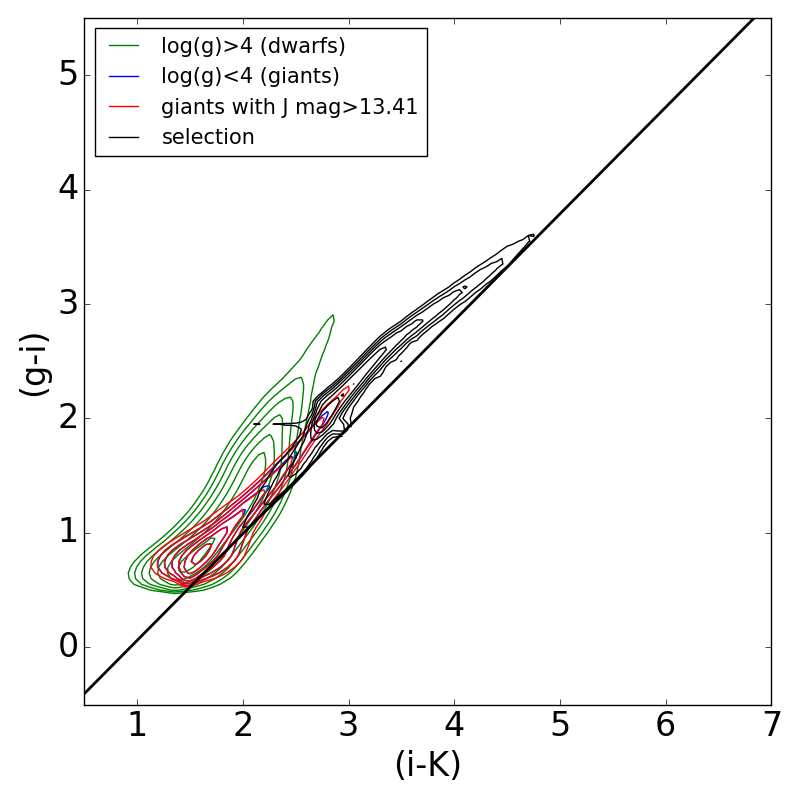}
\end{tabular}
\caption{Selection in color-color space to eliminate quasars, shown for the 11 stacked halos (see Section \ref{subsec:mgiants-present:select}). Stars above the diagonal line are selected. Color scheme is as in Figure \ref{fig:Mgiants_cc}.  Left: the selection operating on the synthetic survey without error convolution. Right: the same selection applied to the error-convolved synthetic survey.}
\label{fig:Mgiants_q}
\end{figure*}

\subsection{Generating the synthetic, error-convolved survey}
\label{subsec:mgiants-present:data}

We use {\sc Galaxia} \citep{2011ApJ...730....3S} to construct a synthetic all-sky survey of the 11 mock stellar halos in \citet{2005ApJ...635..931B}. For this initial effort we do not include foregrounds from the Galactic disk and any smooth halo component, focusing only on the accreted halo. This is somewhat justifiable since we are examining a sample at 100-300 kpc in the halo, but it fails to account for the contamination of the sample by disk dwarfs (\citeauthor{2014AJ....147...76B} estimate this significant contamination to be roughly 80 percent). As done in \citet{2014ApJ...790L...5B}, we restrict the synthetic survey generated by Galaxia to the region of color-magnitude space listed in Table \ref{tbl:tracers}. We generate UBV J, H, and K infrared magnitudes and Sloan g and i visual magnitudes for all stars in this space, as well as sky positions, parallaxes, radial velocities and proper motions.

\begin{table}
\begin{center}
\caption{Constants for photometric error models for ``present-day'' survey (UKIDSS + Sloan)}
\begin{tabular}{cccc}
\hline
Band & $a$ & $b$ & $c$ \\
\hline
Sloan $g$ & $7.5 \times 10^{-7}$ & 0.553 & -0.0361 \\
Sloan $i$ & $2.1\times 10^{-9}$ & 0.814 & 0.0134 \\
UBV $J$ & $8.7\times 10^{-10}$ & 0.9993 & $1.74\times 10^{-4}$ \\
UBV $H$ & $1.6\times 10^{-8}$ & 0.870 & $1.62\times 10^{-5}$\\
UBV $K$ & $2.0\times 10^{-8}$ &  0.888 & $-2.65\times 10^{-4}$\\
\hline
\end{tabular}
\label{tbl:errormodel_present_mg}
\end{center}
\end{table}

We then convolve\footnote{For each star in each coordinate, we calculate the error $\sigma$ from a model based on the ``true'' (unconvolved) values generated by the synthetic survey, then draw a random sample from a gaussian with width $\sigma$ centered on the ``true'' value.} the photometry and proper motions (which are used together to select the M giants) with error models based on present-day surveys. We presume UKIDSS-like photometric errors for the J, H, and K bands and Sloan photometric errors for g and i, with magnitude dependence $\sigma_m$ in all bands  defined by the function
\begin{equation}
\label{eq:em-present-mg}
\sigma_m = a e^{bm} + c,
\end{equation}
where $m$ is the magnitude and $a$, $b$, and $c$ are constants that are different in each band, given in Table \ref{tbl:errormodel_present_mg}. We also presume a constant proper motion error of 2 \unit{mas}{-1} yr based on the SDSS/USNO-B 50-year baseline (Sloan DR10).

\subsection{Selecting candidate M giants}
\label{subsec:mgiants-present:select}

We mimic the series of cuts described in \citet{2014AJ....147...76B} on our synthetic survey, including photometric errors, to select M giants. The effects of each successive cut is illustrated in Figure \ref{fig:prog_cuts} in terms of the distribution of number of dwarfs and giants remaining (left panel) and number of stars remaining as a function of true distance (right panel). Starting with roughly ten million stars from each of the eleven mock halos that fall within the broad color and magnitude ranges in Table \ref{tbl:tracers}, we first choose objects inside the box in IR color-color space shown in Figure \ref{fig:Mgiants_cc}, to conservatively separate giants from dwarf stars. The left-hand panel of Figure \ref{fig:Mgiants_cc} shows the selection operating on the unconvolved synthetic survey, while the right-hand panel shows the selection on the error-convolved data. As shown on the left, this cut would indeed select only giants given perfect data, but as shown on the right, the photometric errors also scatter dwarfs into the box even though the only contamination here is from other stars in the accreted halo. Some evidence of this contamination is seen in the broadening of the black contours indicating the selection in the right-hand panel, relative to the red contours indicating the giants only. Many giants also lie outside the selection box in both the convolved and unconvolved samples. Figure \ref{fig:prog_cuts} shows that with this cut, the number of giants is reduced even more than the number of {\it halo} dwarfs, which would seem to run counter to the intention of this selection. However, in reality the discarded giants would be extremely difficult to separate from foreground {\it disk} dwarfs, which are vastly more numerous than the halo dwarfs considered here.

Another selection is then made in IR-optical space to separate stars from quasars, as shown in Figure \ref{fig:Mgiants_q}. As before, this figure compares the selection made on the unconvolved and error-convoved synthetic surveys in the left- and right-hand panels respectively. There are no quasars in our synthetic survey, but as seen on the left, this cut does still eliminate some real giants as shown in the figure, even when photometric errors are not taken into account. However, the left panel of Figure \ref{fig:prog_cuts} shows that this cut has only a small effect on the target population.

Comparing the numbers of M giants selected by our color cuts made on the unconvolved data versus the same cuts made on error convolved data, we find that convolved data leads to a scattering in and out of selection color space in most cases by a few percent of the number selected for ``perfect'' data.  In 9 of the 11 halos, the net effect of convolution is to scatter stars out of the selection box; in only 2 cases the net effect is to scatter stars in. The percent of the error-convolved and selected sample that is self-contaminating halo dwarfs ($\log g>4$) varies widely across the 11 mock halos, from 4 percent to just over 50 percent, with about 15,000 halo dwarfs on average scattered into our color space selections (ranging from just under 5,000 to nearly 50,000) (see Figure \ref{fig:prog_cuts}). These dwarfs are generally at a heliocentric distance of around 2 kpc, and all are at a distance of less than 6 kpc.

Next, we apply a proper motion selection as in \citet{2014AJ....147...76B}, removing stars with proper motions larger than $2.5\sigma_\mu$ in either RA or Dec, with an average PM error of  $\sigma_\mu = 2$ \unit{mas}{-1} yr as determined for the SDSS-USNOB cross-match used in that work. This cut removes stars with high proper motions, which are assumed to be in the foreground and therefore more likely to be dwarfs than giants. In contrast with \citet{2014AJ....147...76B}, who found this cut to have little effect on their final sample, we find this selection can have a large effect, removing anywhere from 9 up to 50 percent of color-selected, error-convolved stars that from Figure \ref{fig:prog_cuts} appear to be primarily dwarfs within 100 kpc. Once these stars are removed the percent contamination by halo dwarfs decreases dramatically to between 0.06 and 2 percent, with an average among the halos of 250 dwarfs remaining and a range from 40 dwarfs to just under 1000. On average these dwarfs are at a heliocentric distance of around 3 kpc, and all are at distances less than 6 kpc. The proper-motion selection, when combined with the color cuts, is therefore effective at removing the self-contamination of halo dwarfs in the foreground even when the proper motions are not measured very accurately. In the near future, cross-matching WISE, SDSS, and 2MASS observations, and adding PanSTARRS when available, could provide a longer baseline and become even more useful for removing foregrounds (including the halo as well as the disk).

The last step in the selection process is to select stars with simulated, error-convolved $J$ apparent magnitudes greater than 13.41, the $J$ magnitude of the brightest star for which \citet{2014AJ....147...76B} obtained spectra.  This magnitude limit is intended to remove any remaining foreground so we can focus on the truly distant stars, though there is not much of this in our case since we do not include a disk. The right-hand panel of Figure \ref{fig:prog_cuts} shows that this magnitude cut, represented by the purple line, is effective at removing nearby stars. After this selection is made, 64 percent of the remaining stars are more distant than 100 kpc, and nearly 98 percent are more distant than 50 kpc. After the full selection process roughly 10,000 to 300,000 stars remain per mock halo, with a median value of around 50,000 (Figure \ref{fig:prog_cuts}, left panel).


\section{The M-giant view of the distant stellar halo}
Having created 11 mock views of a Milky-Way-like halo in M giants, we investigate in this section exactly what this view can show us about the distant stellar halo. In addition to the number of stars expected at these distances, we consider the variation between halos in the distance distribution of tracers and the types of structures seen on the sky (Section \ref{subsec:mgiants-present:sky-distance}). We also investigate what kinds of accreted structures are preferentially sampled by looking at M giants (Section \ref{subsec:mgiants-present:sats}), how we might search for M giants from the same structure (Section \ref{subsec:mgiants-present:search}), and what information is most useful to untangle different accreted structures that overlap each other on the sky (Section \ref{subsec:mgiants-present:untangling}). To assist in interpreting the eventual results of current M giant searches, we also explore how many different structures we expect to see in a UKIDSS-sized box, and how many stars we expect per structure, as well as the variations between halos and for different locations of the survey (Section \ref{subsec:mgiants-present:obs}).

\begin{figure*}
\includegraphics[width=\textwidth,angle=270,scale=0.7]{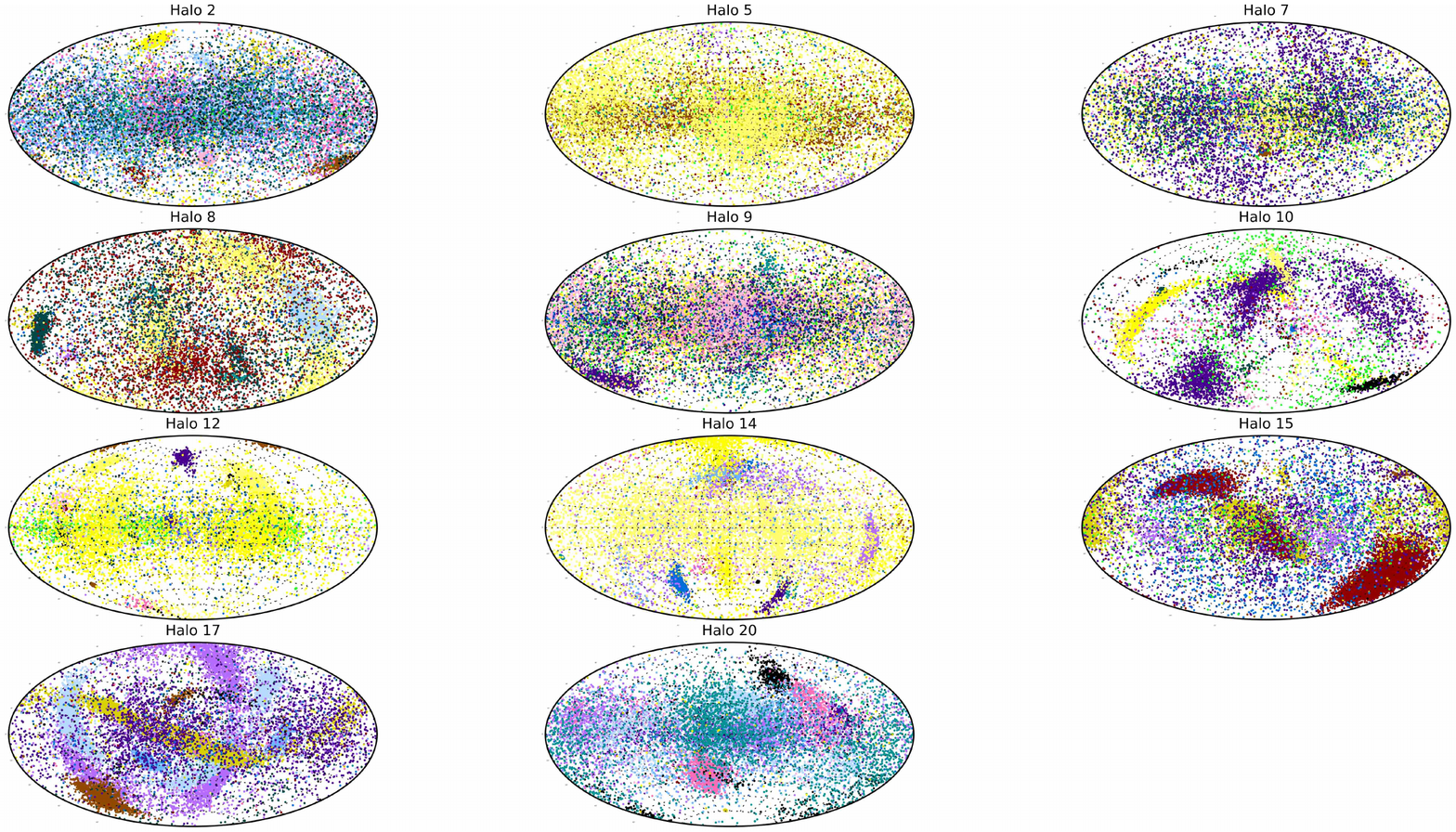}
\caption{All sky map of Mgiants selected as in section \ref{subsec:mgiants-present:select} for each of the eleven mock stellar halos. Different colors represent different accreted satellites. }
\label{fig:Mgiants_allsky_present}
\end{figure*}

\subsection{Sky and distance distributions}
\label{subsec:mgiants-present:sky-distance}
We are studying M giant stars as particularly interesting both because they can be efficiently separated from foreground dwarfs and because  their intrinsic luminosity allows us to use them to trace the outermost reaches of our Galaxy. 
Figure \ref{fig:mg_cumul_present} takes a first look at the full size and extent of the M giant populations in the models by plotting the number of selected M giants as a function of their true distances generated in the Galaxia simulation.
It shows that there are over 1,000 M giants present beyond 100 kpc for all 11 mock halos and over 10,000 for several of the halos. There are also more than 1,000 unbound M giants present beyond 100 kpc for all but one of the mock halos.

\begin{figure*}
\begin{tabular}{cc}
\includegraphics[width=0.5\textwidth]{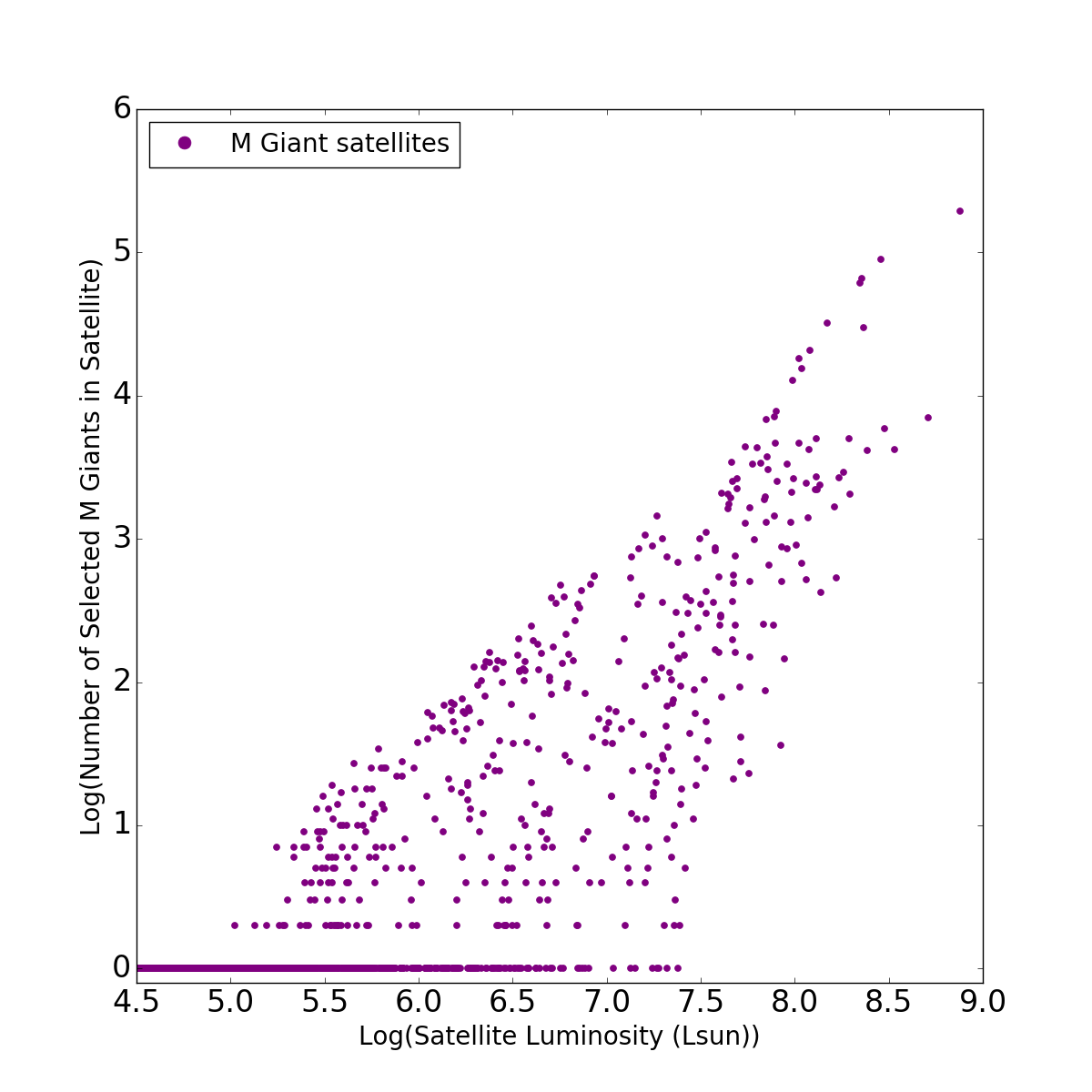} & \includegraphics[width=0.5\textwidth]{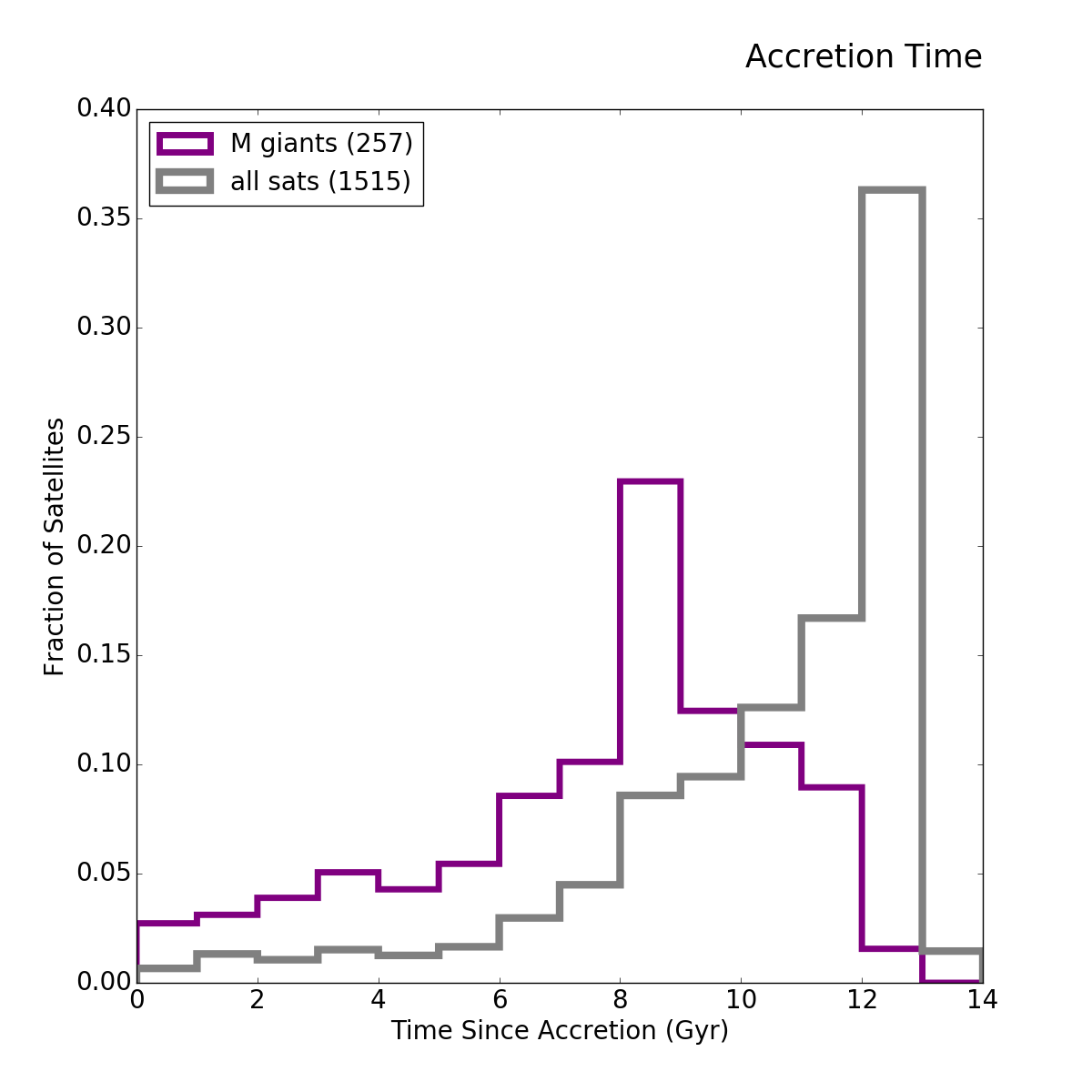} \\
\includegraphics[width=0.5\textwidth]{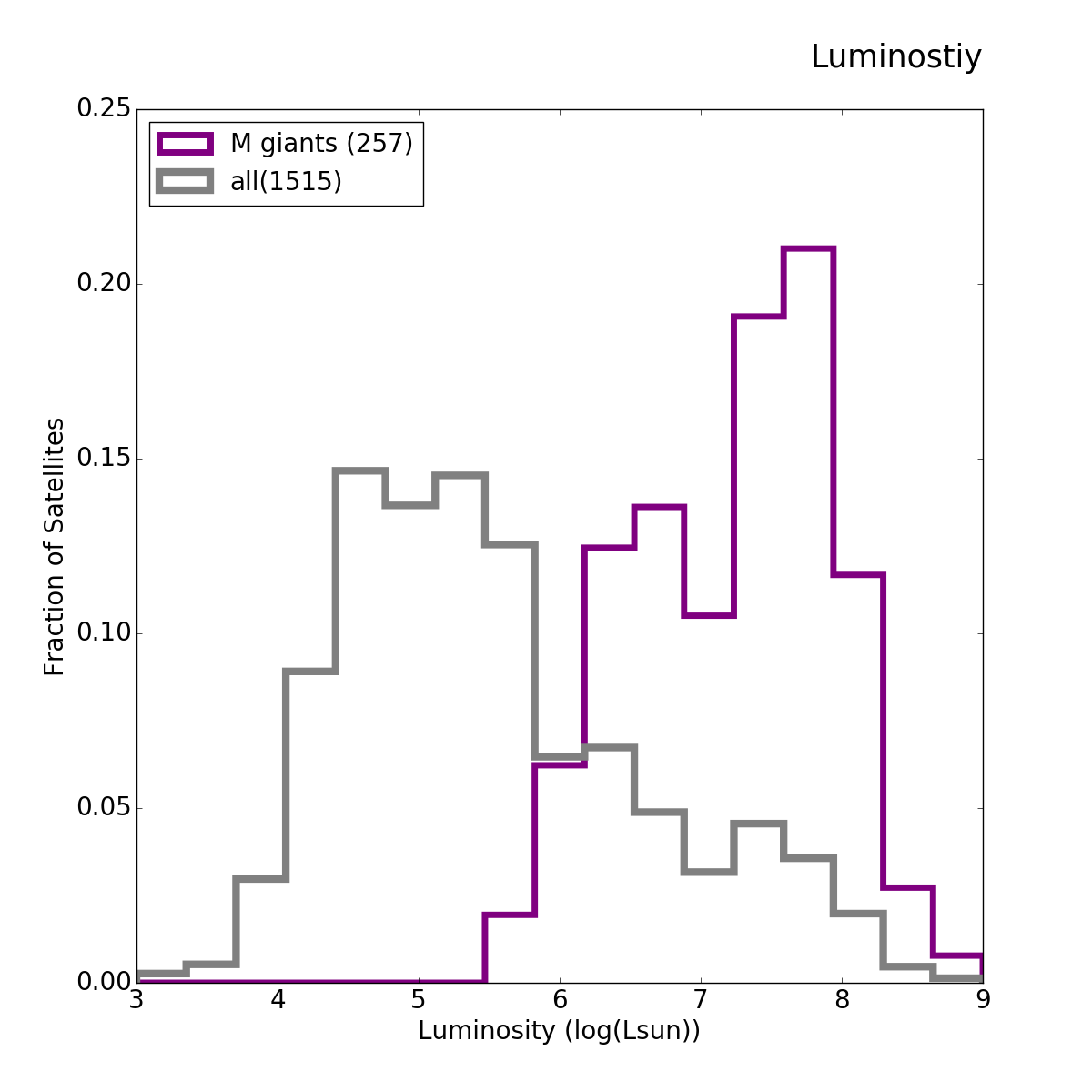} & \includegraphics[width=0.5\textwidth]{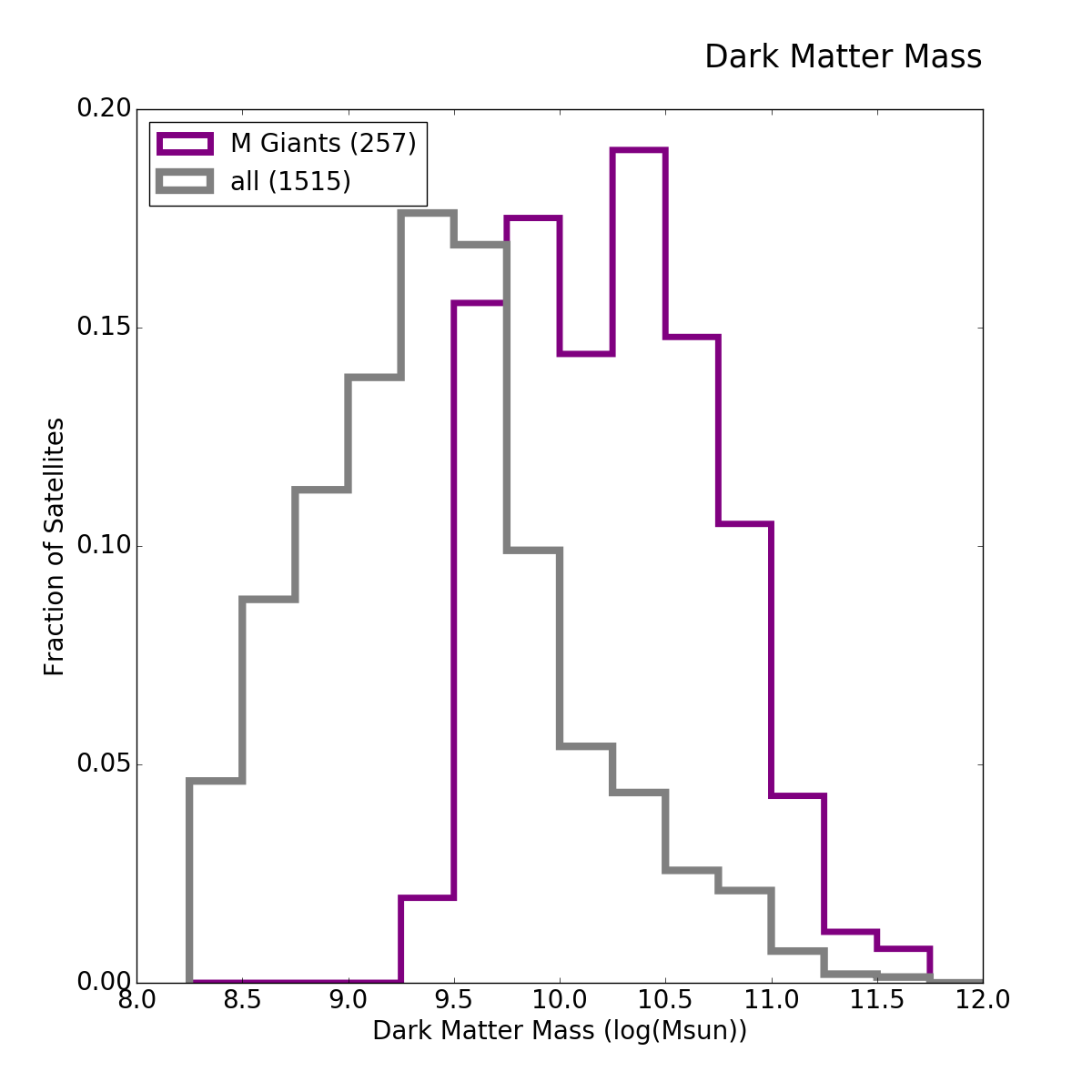}
\end{tabular}
\caption{Top left: Logarithmic distribution of total satellite luminosity versus number of selected M giants in accretion satellite. Other panels: Distribution of various properties of satellites containing 20 or more selected M Giants (purple) compared to the overall satellite distribution (gray). Top right: time since accretion; bottom left: total luminosity; bottom right: total dark matter mass.}
\label{fig:jsat_tacc}
\end{figure*}

When including M giants in bound and unbound satellites in these star counts, as in the left-hand panel, there is over an order of magnitude in variation between the eleven Galaxia halos in the number of selected M giants between 100 and 300 kpc. This variation decreases quite a bit to around 0.5 dex when only unbound stars are plotted as in the right-hand panel. This points to the star counts for some of the simulated halos being dominated by dense, bound satellites within certain distances, whereas other halos have fewer bound satellites in this distance range.  The exception to this would be Halo 10, represented in purple, which appears to have roughly an order of magnitude fewer unbound stars beyond 100 kpc. This is likely because most of the M giants (over 70 percent) in Halo 10 were accreted in a few massive objects about 4 Gyrs ago, which have not yet had time to tidally disrupt. The time of accretion of our selected M giants over all 11 halos peaks at 8 Gyrs ago (see Figure \ref{fig:jsat_tacc}) which allows more structures to become unbound on average than in Halo 10.  

	As expected for a purely accreted population, the eleven mock halos demonstrate a broad variation in the number and type of substructures present at large diatances. Figure \ref{fig:Mgiants_allsky_present} shows all-sky maps in Galactic coordinates of M giants for all 11 mock halos, where each point represents a selected error-convolved M giant color-coded by the satellite it belongs to. In most of the mock halos it is apparent that stars are not evenly distributed across the sky, but there is a good deal of variation. Halo 14, for example, is dominated by one diffuse unbound satellite at central latitudes (represented in yellow), and then several somewhat less diffuse satellites at higher latitudes. Alternatively, Halo 2 has a large diffuse stellar population at central latitudes from many unbound satellites. The M giant map of Halo 12 shows several long distinctive stellar streams (in purple and yellow for example), while Halo 20 provides two particularly good examples of M giants that were accreted as part of the same satellite but are now located in unbound groupings at opposite ends of the halo.  These two large structures (in purple and pink in Halo 20) form over-densities in these projections despite being unbound.
    
    Overall our models suggest that there are around 1,000 to 10,000 unbound M giants beyond 100 kpc, in other words less than one per square degree on average.  However, these stars are distributed non-uniformly across the sky, and this distribution varies significantly between halos: in some halos even the unbound structures are fairly dense and compact, while in others the accreted structure is spread over much broader areas of sky. In addition, associated debris can cover hundreds to thousands of square degrees in structures, sometimes on opposite sides of the sky. 

\subsection{Distribution of M giants among the accreted satellites}
\label{subsec:mgiants-present:sats}

Now we examine the properties of the satellites containing M giants selected by our observational cuts. In the top left panel of Figure \ref{fig:jsat_tacc} we compare the number of M giants in a satellite to the total luminosity of that satellite, and find that for satellites containing more than about 10 M giants, satellite luminosity is strongly correlated with the number of M giants associated to that satellite. However, for satellites containing only a few M giants there is a very large range of luminosities, as much as three orders of magnitude. These trends come from the assumptions in the models that infalling satellites follow a stellar mass-metallicity relation. Since giant stars only become red enough to reach M spectral type if their metallicity is above [Fe/H]$\sim -1$  then the number of M giants can be expected to loosely track the overall luminosity of a satellite, but can not be used to make any precise assumptions about the total luminosity of a satellite. Metallicity measurements for the individual M giants in a structure, however, could help break this degeneracy as well as potentially helping to disentangle different accreted structures from one another, as we will discuss briefly in Section \ref{sec:ratios}. 

	As we pointed out in Section \ref{subsec:mgiants-present:select}, the color selections we made tend to favor metal- and alpha-rich stars. In the remaining three panels of Figure \ref{fig:jsat_tacc} we examine how this affects the types of satellites that are prominent in M giants. We compare the properties of satellites containing a minimum of 20 selected M giants (i.e. satellites that could potentially be identified using M giant observations) to the overall satellite population for all 11 halos combined. We observe that satellites containing 20 or more M giants tend to be more luminous and have a greater dark matter mass, compared to the general population. Our selected satellites also tend to have been accreted more recently than the general population, peaking between 8 and 9 Gyrs ago as opposed to around 12 Gyrs ago (top right panel of Figure \ref{fig:jsat_tacc}). This preference for more recently accreted satellites is consistent with our cosmological model because we are attempting with our absolute magnitude cut to select primarily satellites in the outer halo. It is also related to the preference for higher metallicity satellites established by the color cuts. \citep[See also][for a related discussion.]{2011ApJ...728..106S}

\begin{figure*}
\begin{tabular}{cc}
\includegraphics[width=0.5\textwidth]{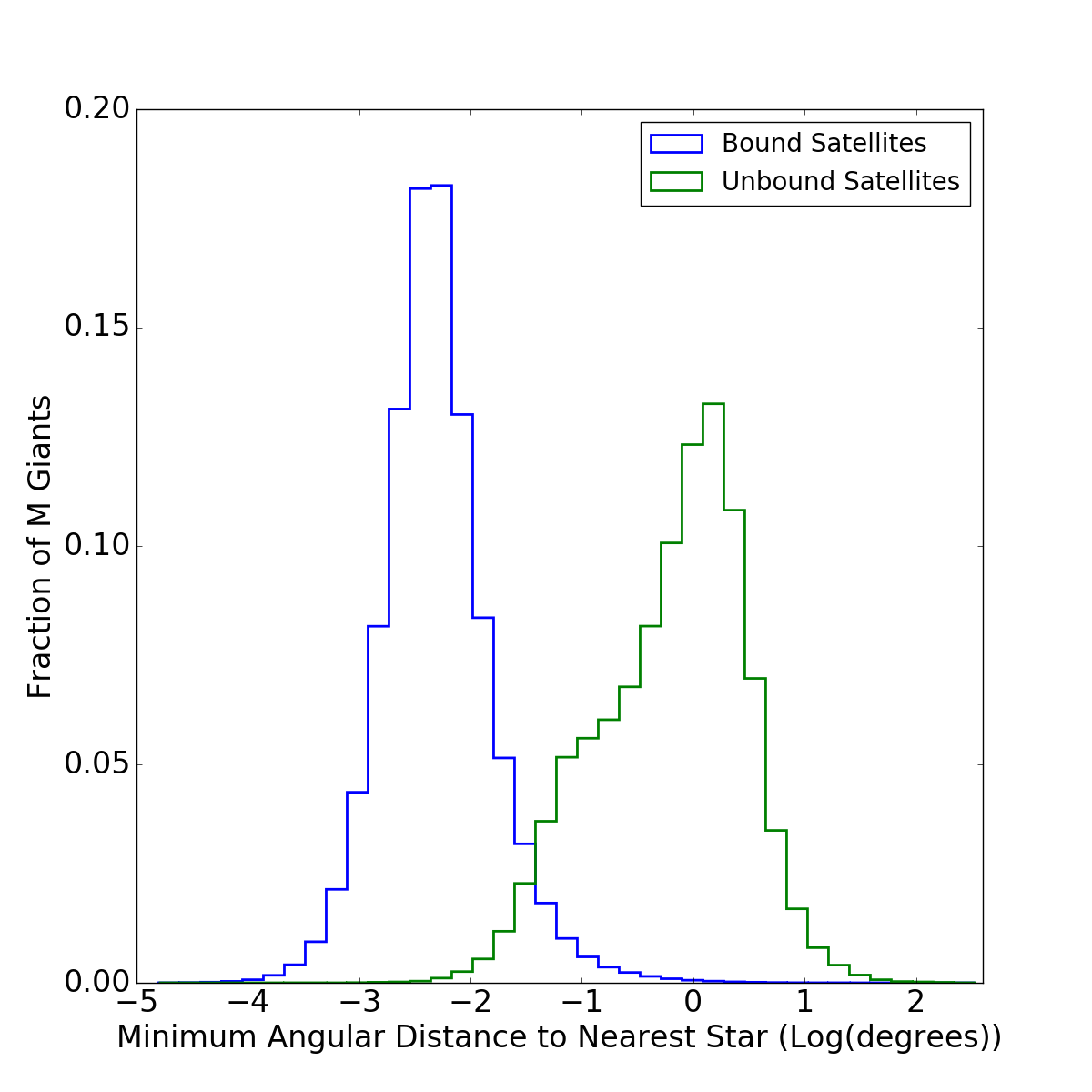} & \includegraphics[width=0.5\textwidth]{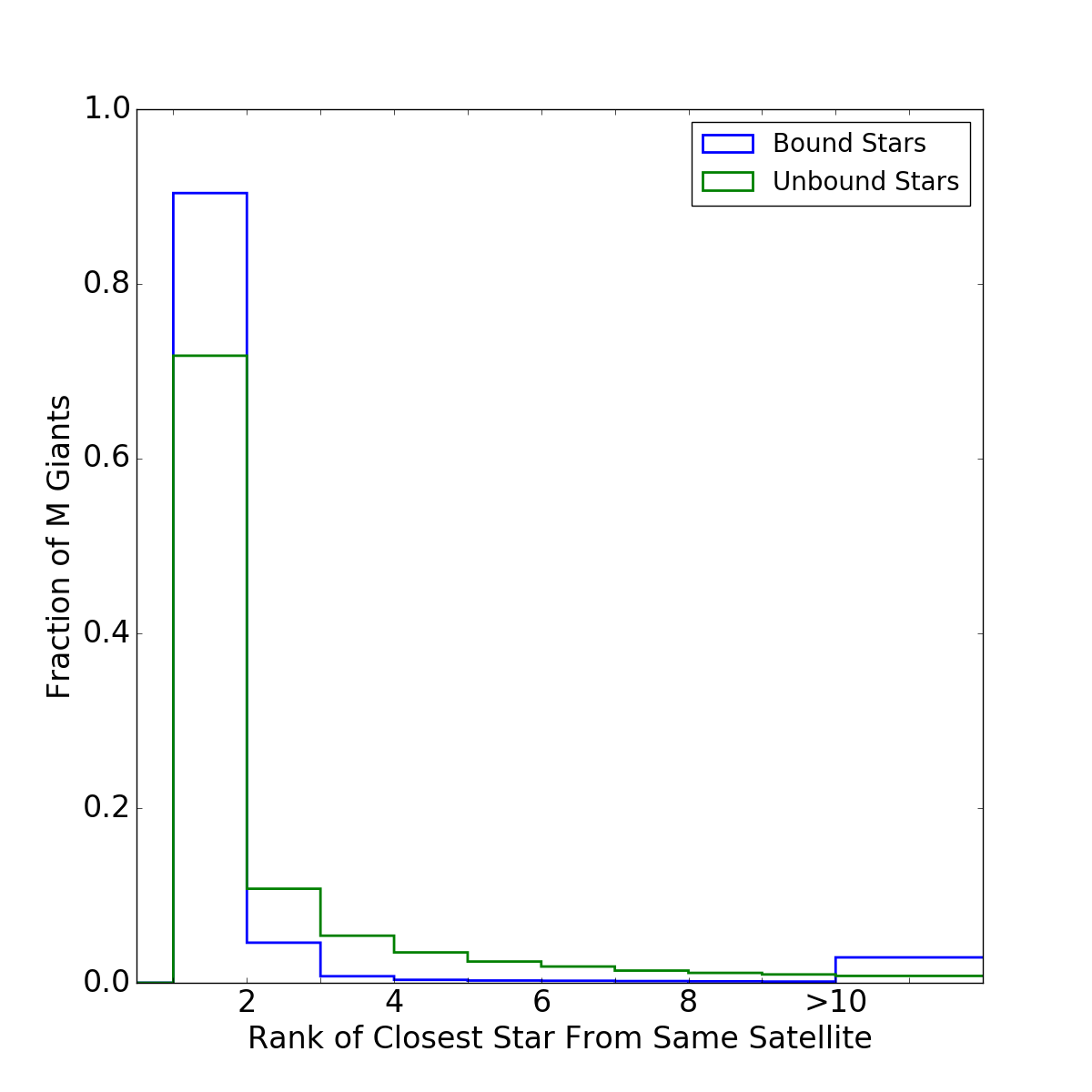}
\end{tabular}
\caption{Left: Distribution of the minimum angular distance (in log(Degrees)) to the nearest M giant that was accreted as a member of the same satellite. Right: Distribution of the rank in distance of the closest M giant in the same satellite: 1 if the closest M giant is in the same satellite, 2 if the second closest M giant is the same satellite but the first closest is not, and so forth.  These distributions are for all selected M giants of all 11 halos. Satellites that are still bound are in blue, satellites that are now unbound are in green.}
\label{fig:nearestM}
\end{figure*}

\subsection{Finding building blocks of the accreted halo}
\label{subsec:mgiants-present:search}
    
    We now consider the prospect of separating the different accreted satellites traced out by the M giants in our selection. One possible approach to searching for structures is simply to search for close pairs of tracers from the same stellar population. \citet{2014ApJ...793..135S} and \citet{2015AJ....150..160B} use this approach to search for new bound satellite galaxies in samples of RR Lyrae, so we investigate whether this is also an effective technique for M giants.  The left panel of Figure \ref{fig:nearestM} shows the distribution of distances to the nearest star of the same satellite for the selected M giants for all 11 halos combined. In the right panel of the figure, we show the distribution of the rank of the nearest M giant that was accreted in the same satellite again for the selected M giants of all halos combined. This rank is 1 if the closest M giant is in the same satellite, 2 if the second closest star is the closest star that is in the same satellite, and so on.  In both panels, the distribution for bound satellites is shown in blue, and for unbound structures in green.  In both scenarios, bound and unbound, the closest M giant to each M giant tends to be from the same satellite.  This chance is just over 70 percent for unbound structures, and around 90 percent for bound satellites.  It is somewhat more likely that the nearest star from the same satellite is the second through ninth closest star to a particular M giant in the unbound case than in the bound case. Less than ten percent of the time will the nearest M giant from the same bound satellite be greater than the second closest star, and the same is true for the fifth closest star for a member of an unbound satellite.  The fraction of stars from the same satellite that are more than ten stars away from the nearest star they were accreted with is also shown. This appears to be more common for bound stars but still occurs less than ten percent of the time.
    
    That stars closest to each other in our data tend to have been accreted from the same satellite, even if they have since become unbound, is a promising development for being able to identify stellar structure.  The left panel of Figure \ref{fig:nearestM} helps to provide a diagnosis when two M giants are discovered near to each other as to whether these M giants are part of a bound or unbound structure. The distribution of distances for bound satellites (in blue) clearly peaks around a few arcseconds, while the distribution for the unbound satellites(in green) clearly peaks around a few degrees, with little overlap.  Therefore, the angular distance between two observed M giants could potentially be used to determine whether they belong to a bound or unbound structure, although doing so would require excellent foreground removal (remember that we do not include the disk foreground in our models).  The distribution of distances also points to the need of contiguous observational fields covering at least a few square degrees to have a good chance of locating two M giants from the same unbound structure.

\begin{figure*}
\includegraphics[width=\textwidth,angle=270,scale=0.7]{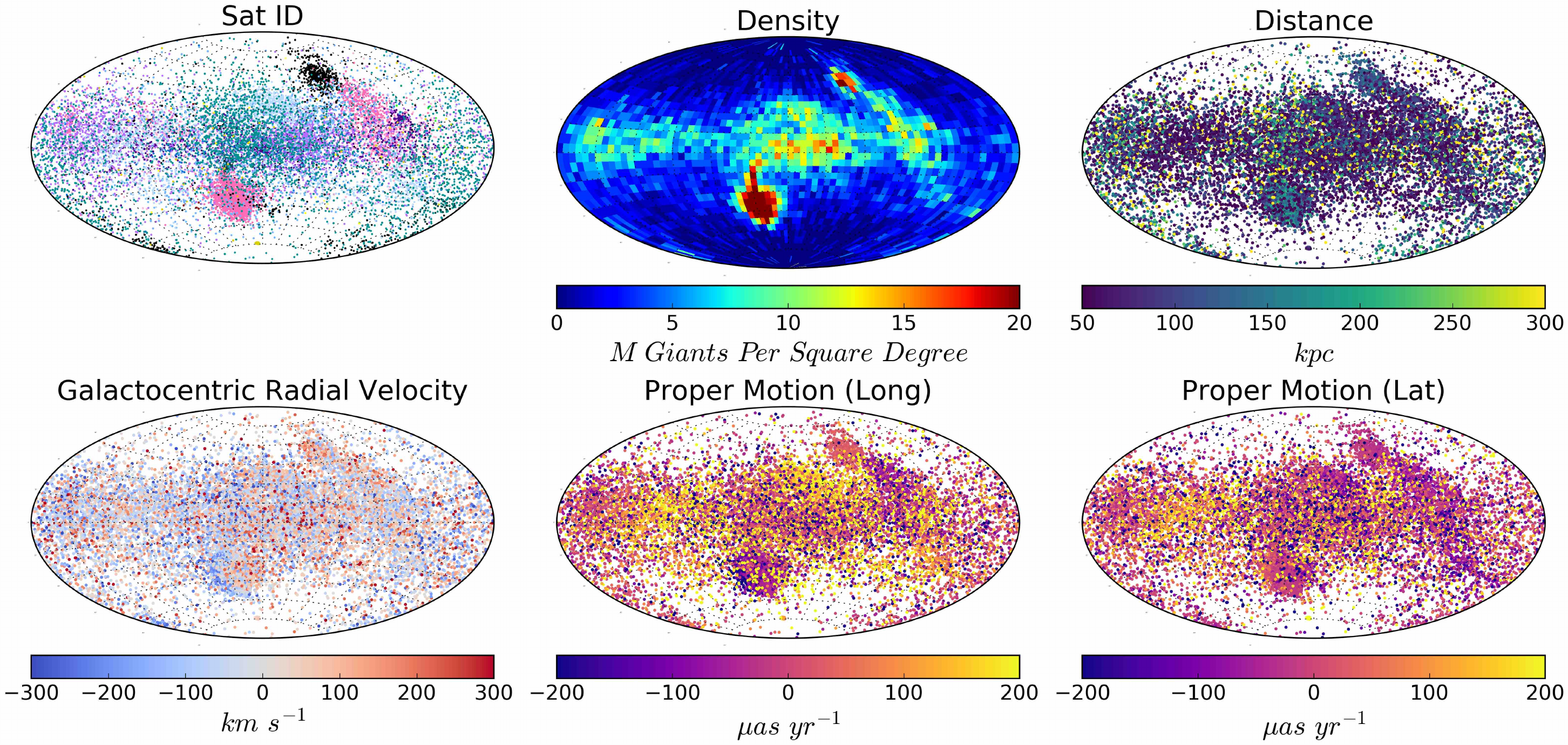}
\caption{Six all-sky views of selected M giants for Halo 20. Each map is color coded by different properties: the satellite the star was accreted with (top left), the density, per square degree, of M giants from unbound satellites (top center), the heliocentric distance with 20 percent errors drawn from a Gaussian (top right), the galactocentric radial velocity of the star (bottom left), and the galactocentric proper motion of the star (longitudinal: bottom center, latitudinal: bottom right).}
\label{fig:4x4M}
\end{figure*}

\subsection{Distinguishing different accreted structures}
\label{subsec:mgiants-present:untangling}
	Finally, we explore which observed quantities are likely to be most useful in determining which stars belong to the same structure. In Figure \ref{fig:4x4M} we view an all sky map of the selected M giants of an example halo, Halo 20, color-coded by six different properties: satellite the star was accreted with; stellar density, binned by square degree and only including unbound satellites; heliocentric distance with 20 percent errors drawn from a Gaussian; galactocentric radial velocity; galactocentric longitudinal proper motion; and galactocentric latitudinal proper motion.  Stellar over-densities (top center) clearly occur at locations where even unbound satellites are present, and in most of the stellar halo there appears to be at least one selected M giant present per square degree, especially at lower galactic latitude.  Close examination of the longitudinal and latitudinal proper motion all-sky maps (bottom center and right) show that both components of a star's proper motion can be used to differentiate between overlapping structures, as stars that are in close proximity to each other in galactic latitude and longitude, but not belonging to the same accreted satellite tend to stand out from each other by having different proper motions. In addition stars from the same accreted satellite tend to have similar proper motions. For example, the satellite that is dark pink in the top left plot has a relatively consistent latitudinal proper motion of around 0 $\mu$as/yr both at its lower latitude and upper latitude positions.  The many non-pink stars (i.e. stars from other accreted satellites) that are seen overlapping the pink satellite in the top left panel also tend to be discernable from the pink satellite when looking at their latitudinal and longitudinal proper motions in the bottom right and bottom center plots.  In the longitudinal proper motion plot (bottom center), stars belonging to the satellites colored in turquoise and light blue in the upper left plot can be distinguished where they overlap the dark pink satellite at lower latitudes on the sky, because of the difference in average latitudinal proper motion. 
    
    Not every star's association can be specified using proper motions alone: different satellites can still have similar proper motions and there is some variation in proper motions among stars belonging to the same satellite. However, for satellites that are distinguishable in the figure, the differences between the average proper motions of different accreted structures seem to be on the order of 50--100 $\mu$as/yr, whereas the spread within structures is often less than this. This indicates that PM measurements with 50--100 $\mu$as/yr uncertainty (per star) could be sufficient to start to distinguish different tidal streams at these distances. This is significantly lower than current PM uncertainties, but Gaia and later LSST and WFIRST are expected to achieve this precision. The orbit distribution we assume for the accreted satellites is fairly radial, so this is a fairly conservative estimate of the variation in proper motions.

\begin{figure*}
\begin{tabular}{cc}
\includegraphics[width=0.5\textwidth]{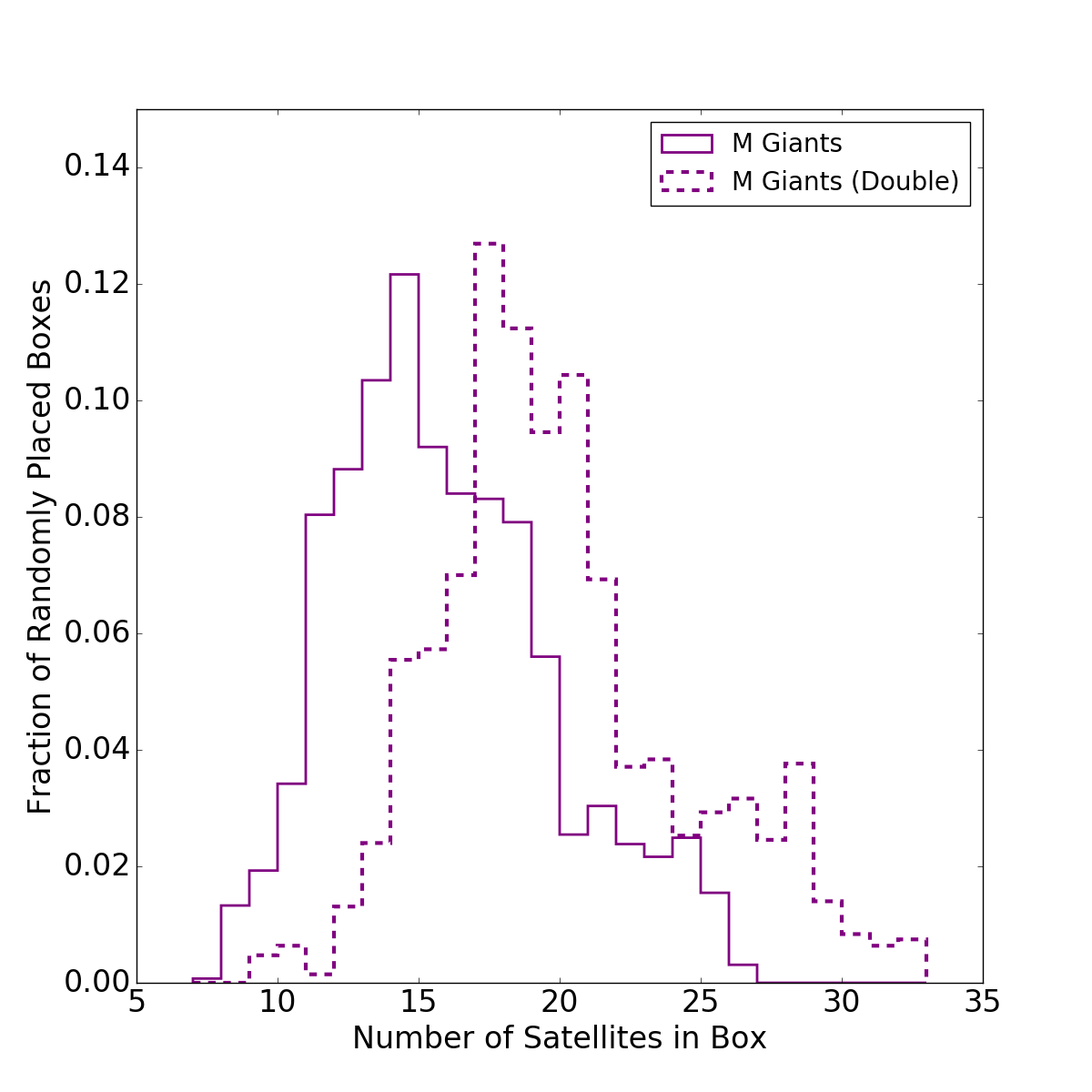} & \includegraphics[width=0.5\textwidth]{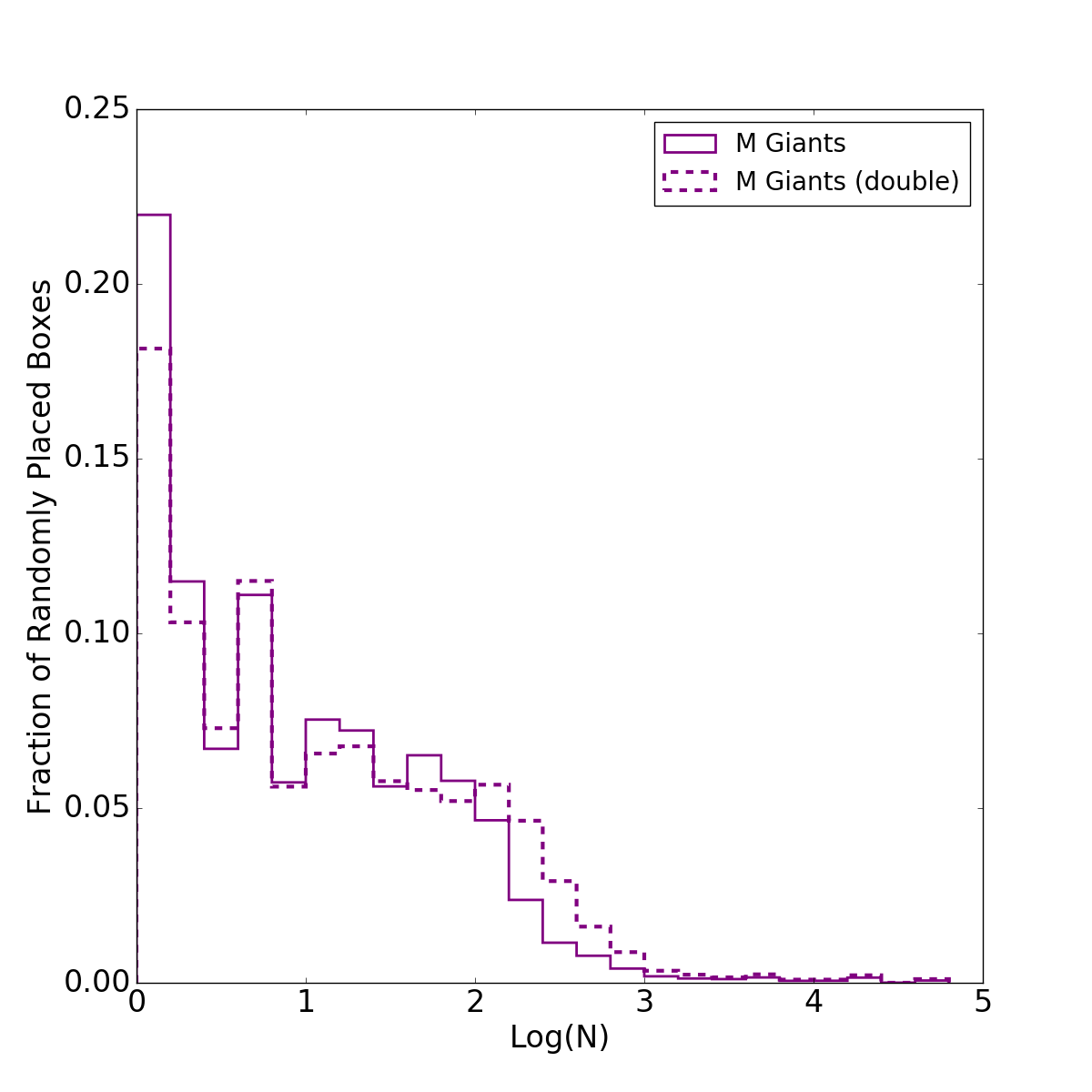}
\end{tabular}
\caption{Left: Distribution of the number of satellites containing M giants in a randomly placed 2200 square degree survey (solid), and a survey twice the size (dashed). Right: Distribution of the number of M giants from the same satellite that fall in a 2200 square degree survey (solid), and survey twice the size (dashed).}
\label{fig:mdub}
\end{figure*}

    Differences between accreted structures are significantly less apparent in radial velocities (bottom left): the overlapping structures seem to have more intrinsic radial velocity variation than proper motion. This is likely because many of the unbound structures at these radii are shells: that is, they are tidal debris in the process of piling up near apocenter. Material truly at apocenter will have zero radial velocity, but this debris is bracketed on one side by outgoing material headed to apocenter and on the other by infalling debris that has already turned around, producing a distinguishable gradient in RV over a single structure. For example, the dark pink satellite that was somewhat uniformly colored in the both proper motion plots has nearly the full range of radial velocities. This makes it very difficult to pick out the overlapping satellites, although the occasional overlapping star from a different satellite still does pop out, such as the two particularly red (high positive radial velocity) stars in this region, which are members not of the pink satellite but of the one color-coded as turquoise in the upper left panel.

    Although RVs are perhaps less useful for disentangling different structures at these distances in the halo, they are very useful for constraining the Milky Way mass profile, since the gradient in RV apparent for a few structures in Figure \ref{fig:4x4M} depends directly on the local radial acceleration \citep{2013MNRAS.435..378S,1998MNRAS.297.1292M}. Errors in the heliocentric distances of the selected M giants, as shown in the top right panel of Figure \ref{fig:4x4M}, make distance less useful for differentiating structures, even many of the unbound structures in Halo 20 are shells, where most of the material is at apocenter. Shells tend to be only a few kpc thick, far less than the distance errors in our model, which makes them difficult to distinguish. Instead, M giants that are members of the same accreted satellite, whether or not they are still bound, seem more likely to be more closely located in terms of latitude and longitude than in terms of radial distance. We conclude that if the Milky-Way accretion history resembles that of Halo 20, we can find unbound structures on the sky as over-densities, potentially disentangle overlaps using proper motions, and then use their RVs to constrain the MW mass profile. For this analysis we have so far considered a structure "identifiable" if it has more than 20 M giants, and we expect that this should be enough for a measurement of the enclosed mass inside a shell, although we will defer testing that assumption to future work.

\subsection{Expectation for and Comparison with UKIDSS Survey}
\label{subsec:mgiants-present:obs}
We now examine the M giant view of the distant stellar halo in the context of the current UKIDSS survey, which subtends about 2200 square degrees of contiguous area on the sky. In our mock halos, the Sun could be located anywhere on the Solar circle (presumed at 8 kpc), so we randomly placed 500 UKIDSS-sized surveys at mid-latitude locations (between either 20 and 70, or $-20$ and $-70$ degrees latitude) in each of the 11 mock halos, for a total of 5500 randomly oriented survey fields. Strictly speaking this is not geometrically equivalent to moving the Sun's location, but since we are focusing on distances beyond 50 kpc the differences are slight. When done this way we can also use the same sample to explore the effect of the location of the survey footprint. The solid lines in Figure \ref{fig:mdub} show the distribution of the number of satellites containing selected M giants located in a survey this size (left), and how many M giants per satellite we would find in the survey (right). Based on our 11 mock halos we find that a roughly 2200-square-degree survey at mid-latitudes will always consist of multiple satellites, with a median of about 14 different building blocks represented. Most of these satellites (about 80 percent) have more than one M giant star in the survey area; about half of them have more than 10 stars and a few (about 10 percent) have more than 100. We therefore expect that a typical survey of this size will be dominated by one or two structures, but will actually contain stars from a good sample of the halo building blocks.

It is interesting to ask whether increasing the survey size has a significant effect. The dashed lines in Figure \ref{fig:mdub} show the number of satellites and number of stars per satellite change if we double the contiguous survey area to 4400 deg${}^2$. When expanding the survey area to twice the size, the most likely number of satellites contributing increases somewhat, from 14 to 18, but the number of stars per satellite is not greatly affected except at the high end of the distribution (above $\sim 100$ stars). Doubling the sky coverage thus helps map out more completely the larger unbound, but still spatially associated structures, such as the dark red structure in Halo 15 in figure \ref{fig:Mgiants_allsky_present}; as well as stream-like structures, such as those in Halo 17 in figure \ref{fig:Mgiants_allsky_present}, and widely spread diffuse structures, such as the yellow structures in Halo 5 and Halo 14. Interestingly, the number of different satellites recovered in the doubled survey area is increased by less than a factor of two. This suggests that, in general, satellites in our simulations are not evenly spread across the sky, which is supported by the all-sky maps of Figure \ref{fig:Mgiants_allsky_present}.

\begin{figure}
\includegraphics[width=0.5\textwidth]{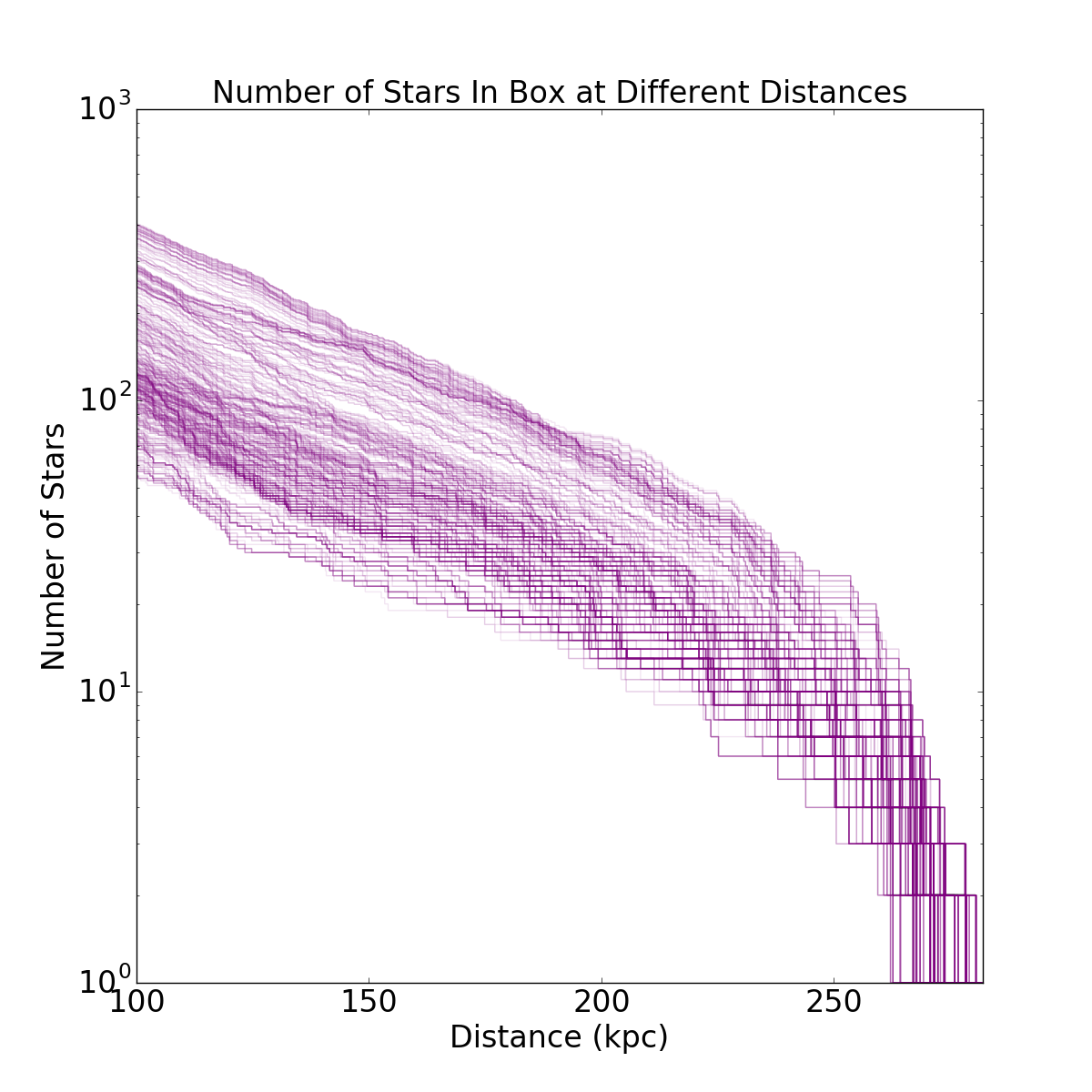}
\caption{Number of M giants (cumulative from the outside in) in 500 randomly placed 2200 square degree fields as a function of true distance, for Halo 5. \label{fig:mg_cumul_present_box}}
\end{figure}

    It is also interesting to ask how representative a given survey field of this size is of the global halo population. In Figure \ref{fig:mg_cumul_present_box} we show the effect of the location of the survey footprint on number of selected M giants of heliocentric distance for Halo 5, represented in orange in Figure \ref{fig:mg_cumul_present}. Halo 5 was chosen because it does not have many bound structures, which we assume will be identified with confidence. We find that changing the location of the survey footprint produces less than an order of magnitude difference in the number of stars detected within a certain distance. The location-dependent spread is also comparable to (or perhaps even less than) the variance in the number of M giants between different halos in the set. We conclude that a survey of a few thousand square degrees at mid galactic latitudes should provide a fairly representative sample of the distant stellar halo. We do note that the primary variation of our field placement is with longitude (although we do consider fields both above and below the Galactic disk), and the Galactic mass model used here is axisymmetric, so any effects of halo triaxiality or filamentary accretion are not present in our mock halos and could affect this result to some degree. However the degree to which the Milky Way halo is triaxial, as well as the directions of any local filamentary flows, are both only poorly constrained; in the absence of better information we consider the axisymmetric assumption to be a reasonable one.

Presuming that our mock halos resemble the Milky Way's, the results of this section demonstrate that a survey field of a few thousand square degrees is both sufficient to get a general sense of the contents of the distant halo and is likely to contain well-sampled debris from about 1--3 accreted satellites that could be associated using additional dimensions of information (beyond sky positions). This bodes well for current searches for distant M giants based on UKIDSS like that described in \citet{2014AJ....147...76B}; it means that the discovery of stars like the one featured in \citet{2014ApJ...790L...5B}, at distances beyond 200 kpc, is expected, and that this star and others like it will likely turn out to be in the most well-sampled tidal structures. Finding even one well-populated accreted structure at large galactocentric distance would be a powerful probe of the Galactic potential.  As the search continues, proximity of these very distant stars on the sky might provide one path to associating them (as we discussed in Section \ref{subsec:mgiants-present:search}). In the future, deeper surveys of a similar size (like the WFIRST High-Latitude Survey) will push further, beyond the edges of the Milky Way.

\begin{figure*}
\begin{tabular}{cc}
\includegraphics[width=0.5\textwidth]{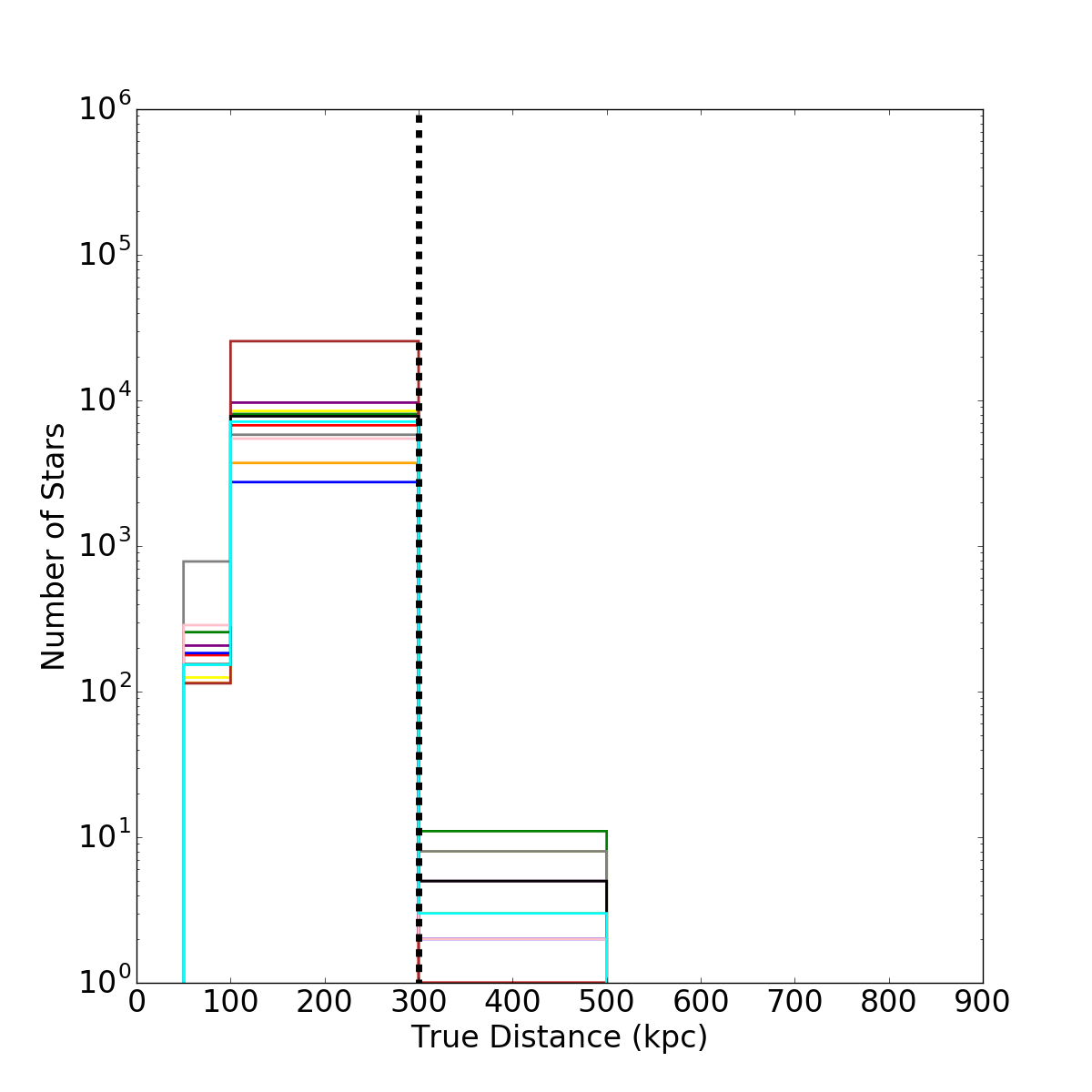} & \includegraphics[width=0.5\textwidth]{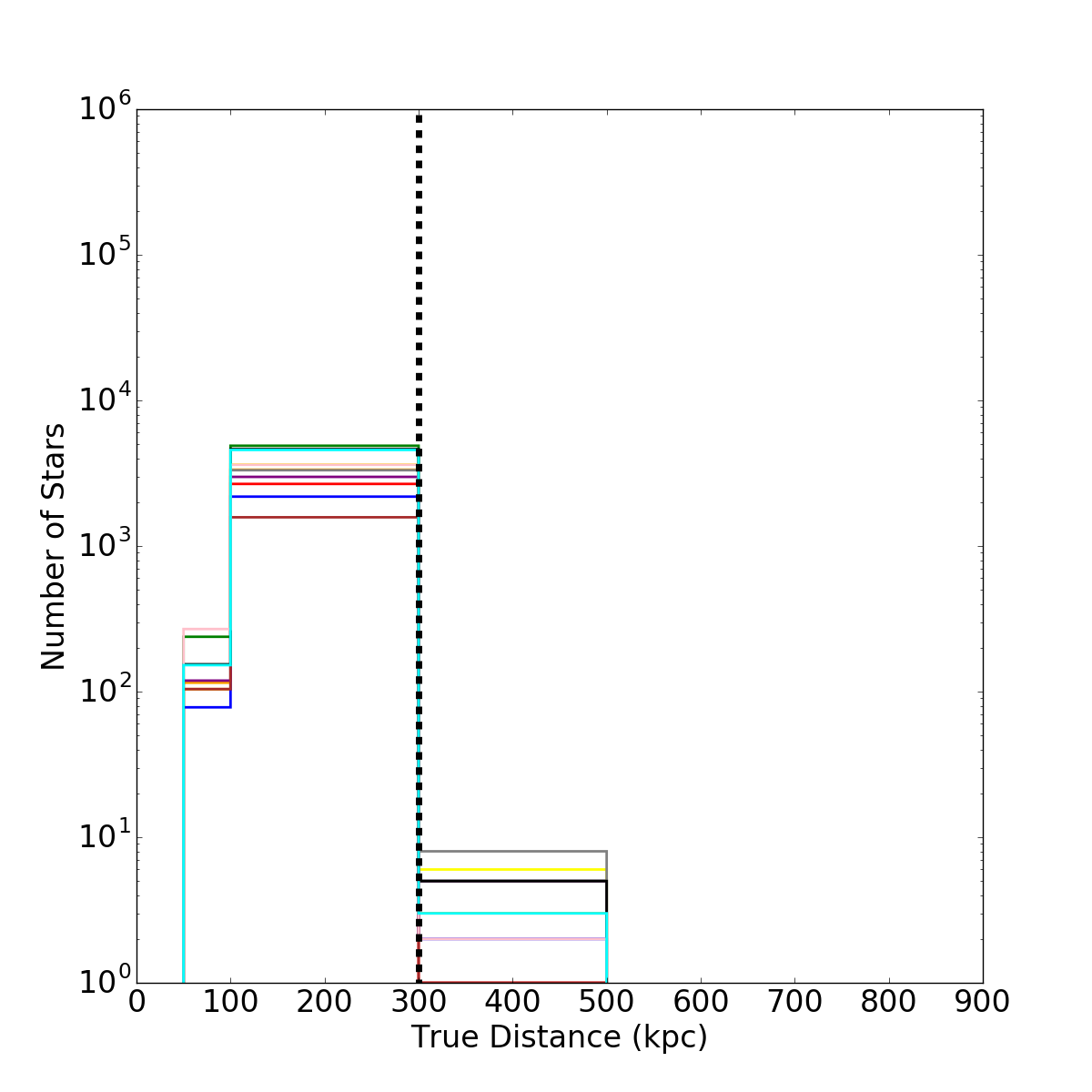}\\
\includegraphics[width=0.5\textwidth]{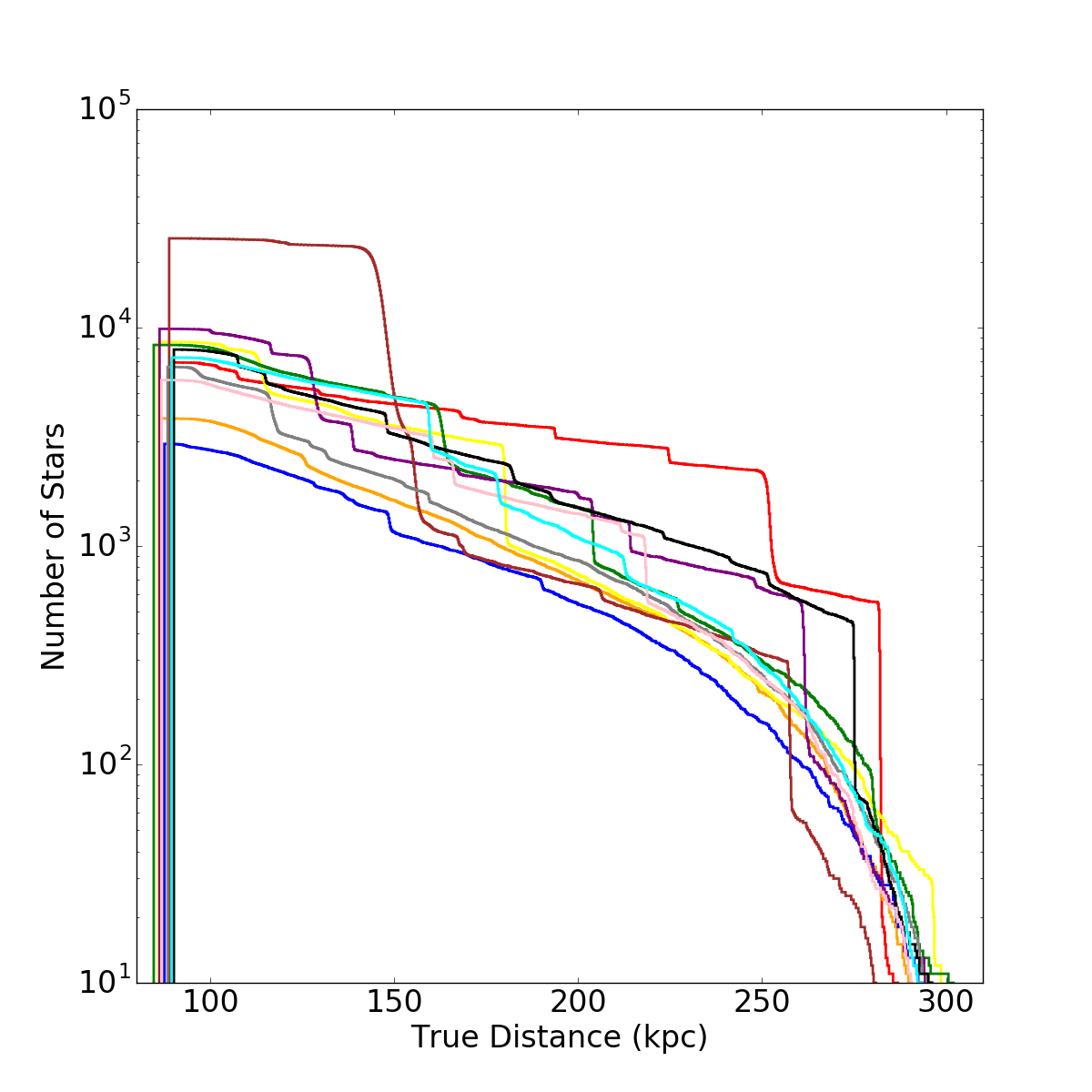} & \includegraphics[width=0.5\textwidth]{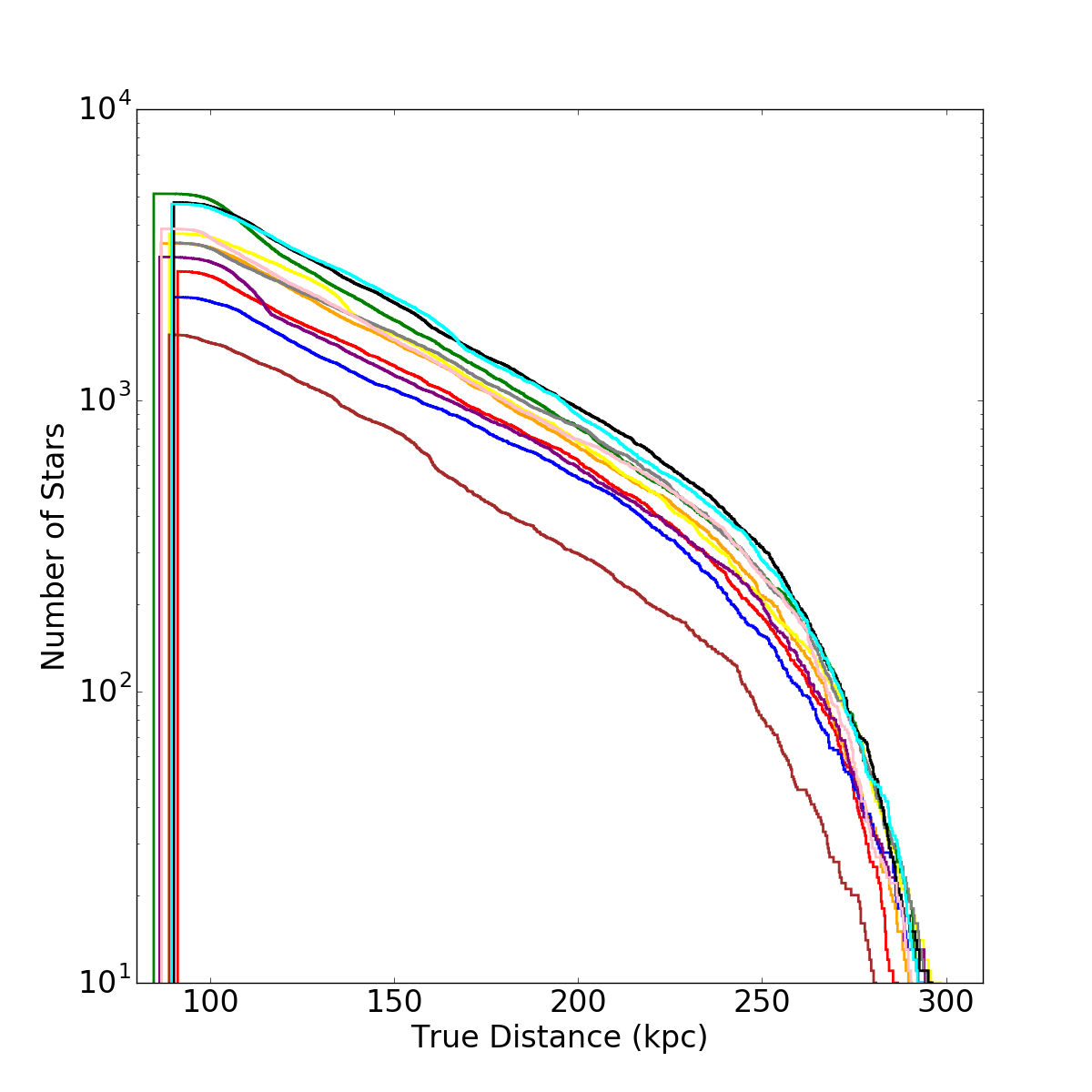}
\end{tabular}
\caption{Number of RR Lyrae in each of the mock halos between error-convolved distances of 100 and 282 kpc as a function of true distance in all substructures (top left) and excluding bound satellites (top right). Cumulative, from the outside in, amount of RR Lyrae between 100 and 282 kpc in error-convolved distances beyond each true distance as a function of true distance for all substructures (bottom left) and excluding bound satellites(bottom right)}
\label{fig:rrl_cumul_present}
\end{figure*}

\begin{figure*}
\includegraphics[width=\textwidth,angle=270,scale=0.7]{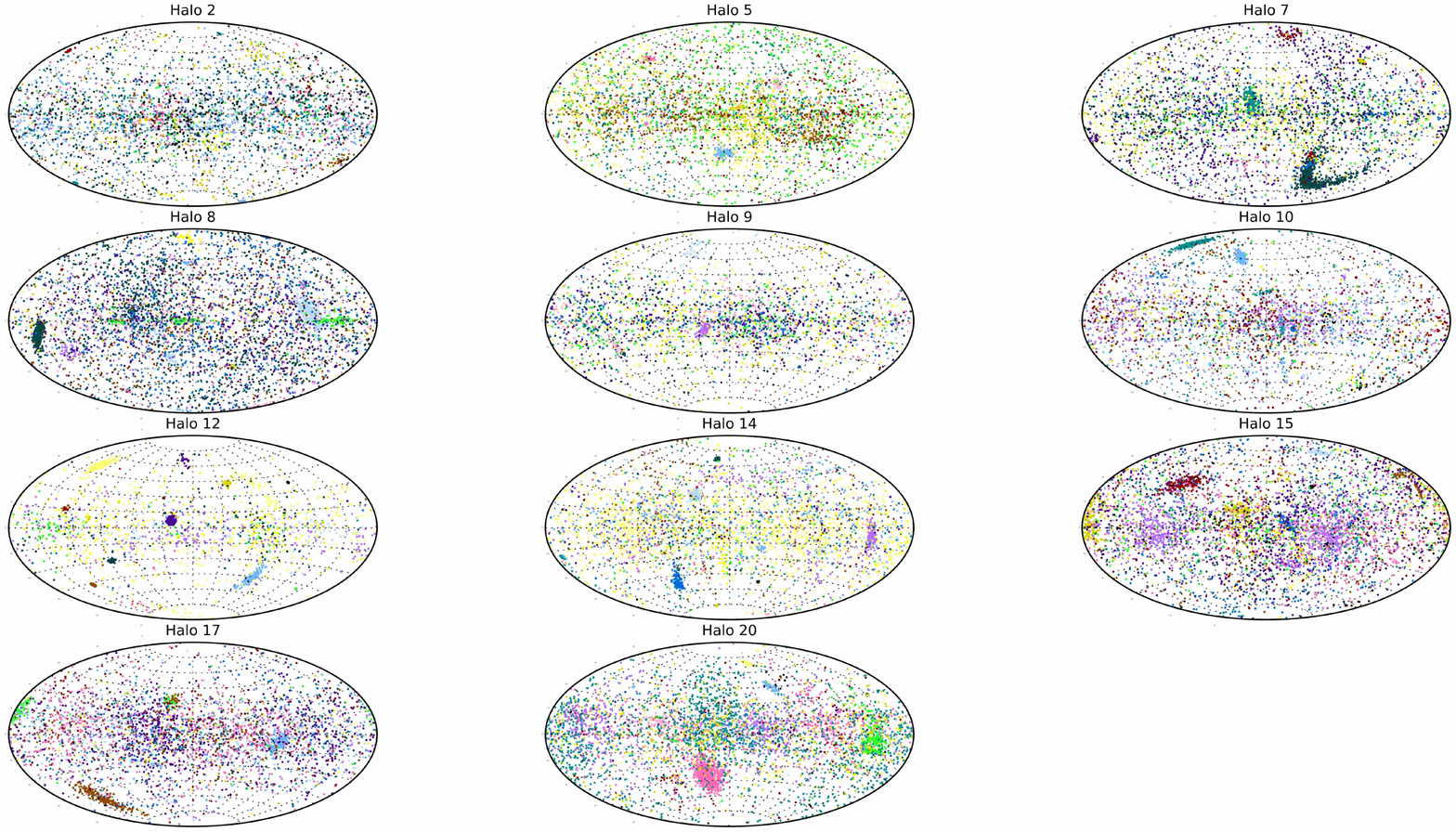}
\caption{All sky map of RR Lyrae between error-convolved distances between 100 and 282 kpc for each of the eleven stellar halos. Different colors represent different accreted satellites.}
\label{fig:rrl_asp}
\end{figure*}

\section{Distant RR Lyrae: Prospects for future surveys}
\label{sec:rrl}
Another stellar tracer in the distant halo that is of great interest is the population of RR Lyrae (RRLe). These tracers are less luminous and less numerous than M giants but have several distinct advantages. First, they are fairly unambiguous to identify from their lightcurves if multiple epochs of observations are available. RRLe also follow a period-luminosity relation that can be used to obtain very accurate distances: to around 5\% in the optical \citep{2003AJ....125.1309C,2010ApJ...708..717S} and less than 2\% in the IR \citep{2016arXiv160401788B}. Finally, they explore different types of objects than M-giant stars since they are only present in populations that contain stars of metallicites [Fe/H]$\lesssim 1$. This makes them powerful tracers of the MW accretion history that are complimentary to M gaints: these very accurate distances can both help to untangle different accreted satellites (in particular, the smaller dwarfs) and better constrain the mass distribution.

In this section we consider prospects for viewing RR Lyrae in the outer halo based on projections for LSST. \citet{2015ApJ...812...18V} have demonstrated that RRLe can be perfectly identified over the entire LSST footprint down to magnitude 24.5 after a few years of operation. Since LSST's footprint spans more than 3/4 of the sky, we use Galaxia to generate a synthetic all-sky survey containing only RRLe at 100 kpc and beyond. Exquisitely accurate (2\%) distances can be obtained for RRLe by linking the optical lightcurves to infrared luminosities, where there is less scatter in the period-luminosity relation, so we also consider what is visible in a survey footprint the size of the High-Latitude Survey (HLS) planned for WFIRST, which will observe a contiguous area of about 2200 square degrees in the infrared down to the H band. The idea would then be to identify and fit lightcurves to RRLe in the HLS field using LSST, then use WFIRST's infrared observations to obtain accurate distances. We tried simulated distance errors of 2 and 5 percent and found few significant differences in the overall samples, so we have conservatively assumed 5 percent distance errors. As with M giants above our primary interest is in the distant halo and so, using the error-convolved distances, we select only RR Lyrae between 100 kpc and the virial radius of the mock halo (282 kpc).

\subsection{The RR Lyrae view of the distant stellar halo}
\label{subsec:rrl:view}

	As for M giants, we first looked at the number of selected RR Lyrae as a function of true heliocentric distance (top panel of Figure \ref{fig:rrl_cumul_present}). Because the distance errors are small for RR Lyrae, most of the selected stars fall within 100 and 300 kpc, with less than a thousand falling above or below this range for each mock halo. All but one of the mock halos has between 1,000 and 10,000 RR Lyrae between this distance range, with one halo having just over 10,000 stars, and all having between 1,000 and 10,000 unbound RR Lyrae in this distance range. In addition, as with M giants, removing RR Lyrae from bound structures leads to greater resemblance in the distance distributions of the different mock halos. 
    
    From comparison with Figure \ref{fig:mg_cumul_present} there appear to be more bound selected M giants than bound selected RR Lyrae in the true distance range of 100 to 300 kpc, by an order of magnitude for several halos. However, there are similar numbers of unbound stars in both tracers in this distance range. This suggests that the M giant view of the stellar halo tends to be more dominated by bound stars. It is apparent that at least an order of magnitude more selected M giants fall below this distance range, because the variation in absolute magnitude of M giants makes an apparent magnitude limit imperfect at eliminating all nearby giants, especially compared to the small distance errors involved in RR Lyrae observation.

\begin{figure*}
\begin{tabular}{cc}
\includegraphics[width=0.5\textwidth]{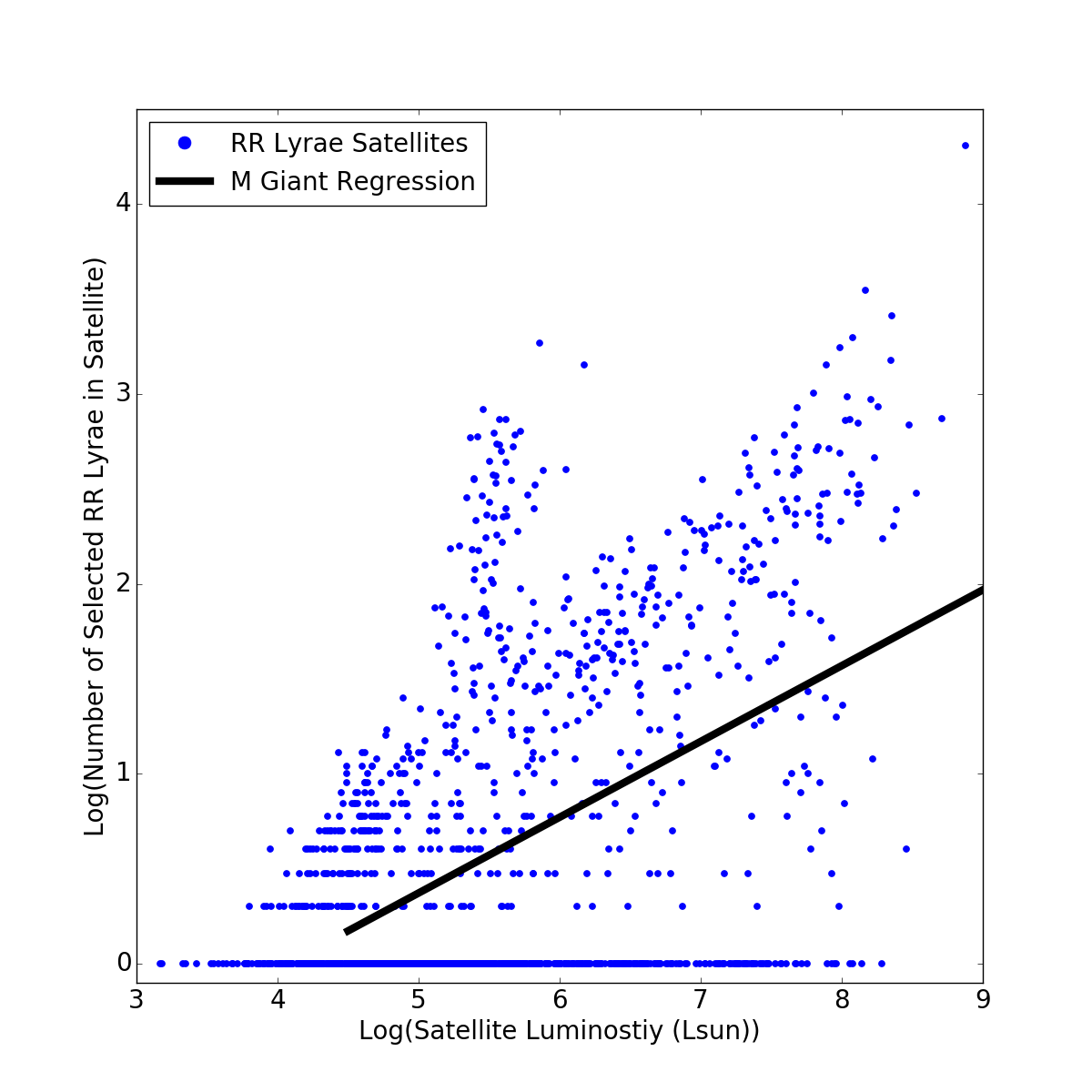} & \includegraphics[width=0.5\textwidth]{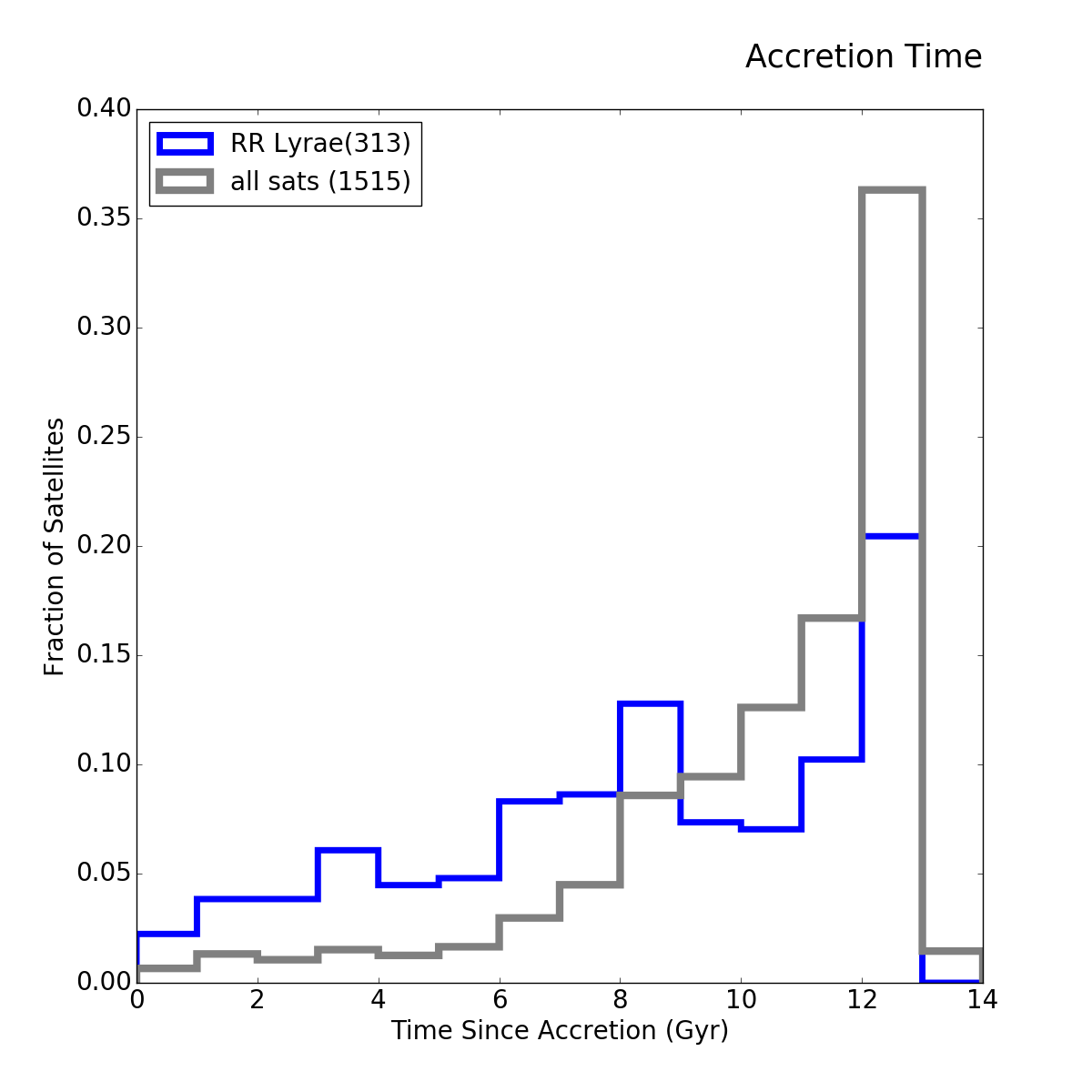} \\
\includegraphics[width=0.5\textwidth]{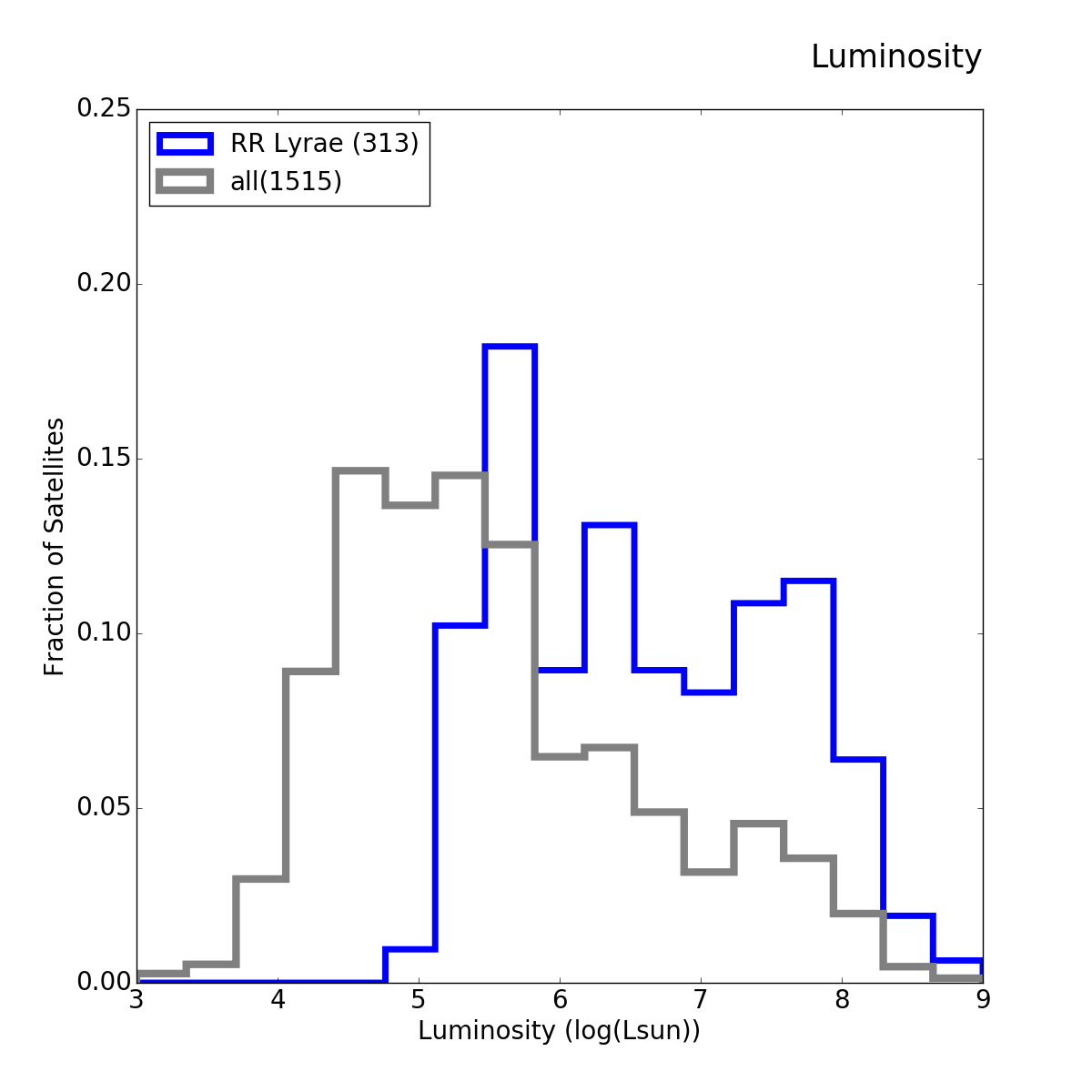} & \includegraphics[width=0.5\textwidth]{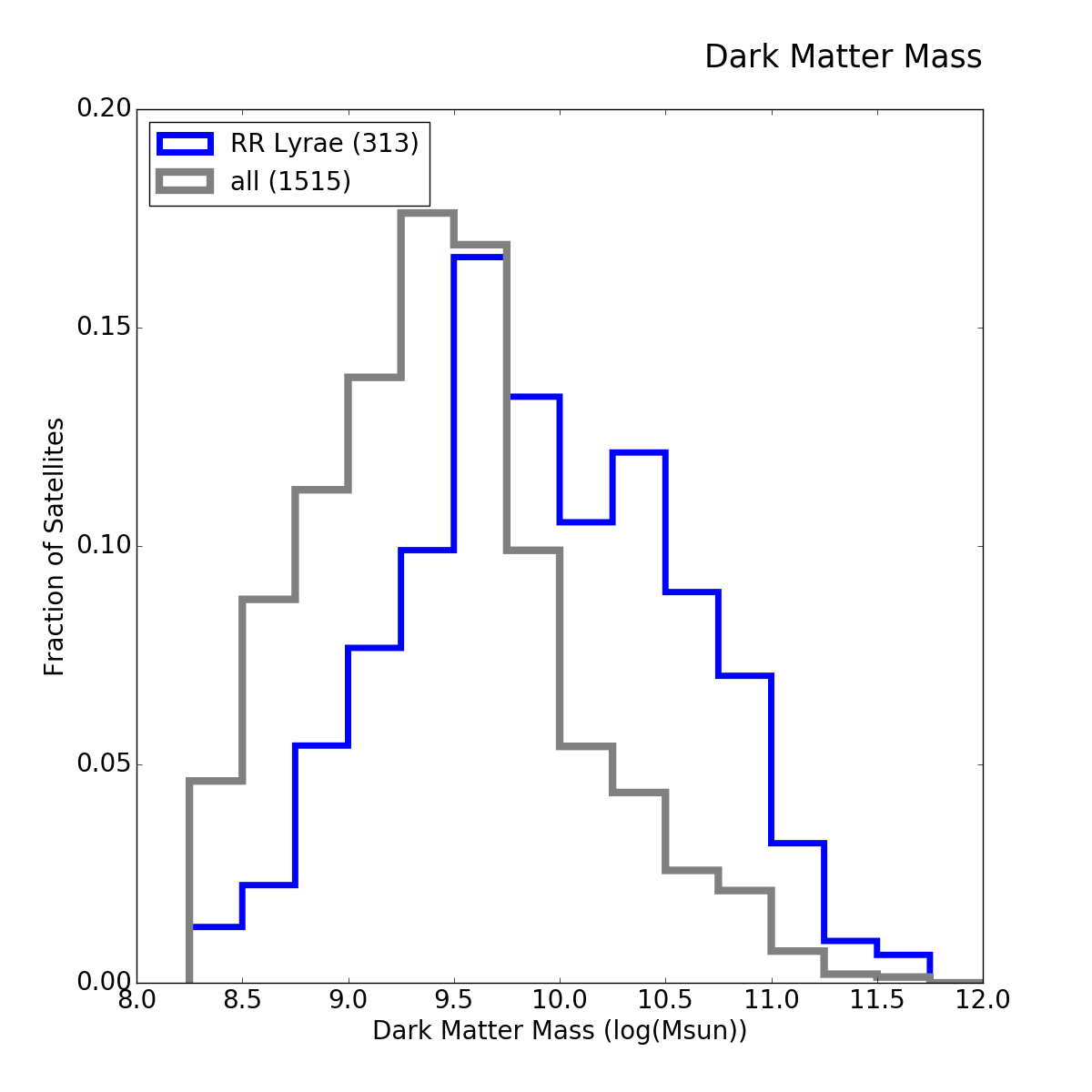}
\end{tabular}
\caption{Top left: Logarithmic distribution of  total satellite luminosity versus number of selected RR Lyrae in accretion satellite. The black line represents a linear fit to the corresponding M giant distribution. Other panels: Distribution of various properties of satellites containing 20 or more selected RR Lyrae (blue) compared to the overall satellite distribution (gray). Top right: time since accretion; bottom left: total luminosity; bottom right: total dark matter mass.}
\label{fig:rrl_prop}
\end{figure*}

    Thanks to the exquisite quality possible for RR Lyrae distance measurements, we can also consider the cumulative distance distribution of selected RR Lyrae as a function of true heliocentric distances (bottom panel of Figure \ref{fig:rrl_cumul_present}), both for all substructures (left) and excluding bound satellites (right). Beyond 100 kpc for 10 out of the 11 halos there are between 2,000 and 10,000 total selected M giants, and between 1,000 and 10,000 unbound M giants, with Halo 12, represented in brown, being a bit of an outlier with a greater amount of bound stars and fewer unbound stars. As for M giants the cumulative distance distribution becomes very consistent among the different halos when excluding stars in bound satellites (with the aforementioned exception of Halo 12). In both the M giant and RR Lyrae tracer we can overall draw the conclusion that if our models are representative of the Milky Way, there are likely thousands of observable stars beyond 100 kpc, and that the distance distribution of tracers from unbound structures is fairly similar across all the mock halos.

	Maps of the RR Lyrae distribution also show a wide variety of bound and unbound structures, but differ in a few ways from what we saw for the M giants in the previous section. Figure \ref{fig:rrl_asp} shows the all-sky maps in Galactic coordinates of RR Lyrae with error-convolved distances between 100 and 282 kpc for all 11 Galaxia halos (again, color-coded by satellite). The halos are more sparsely populated than the M giant maps, both due to there being a larger population of M giants and to the apparent magnitude cuts made for the M giants that do not eliminate as many stars outside the more specific distance range set for the RR Lyrae. The greater number of M giants represented in Figure \ref{fig:Mgiants_allsky_present} than in Figure \ref{fig:rrl_asp} leads to several structures being more clearly defined in the M giant tracer than in the RR Lyrae tracer, such as the large red structures in Halo 15 which barely appear in the RR Lyrae tracer but which are extremely prominent in the M giant tracer.  The looser distance cuts in the M giant tracer also lead to more stream-like structures being picked out in Figure \ref{fig:Mgiants_allsky_present} than in Figure \ref{fig:rrl_asp} such as those in Halo 17 and those represented in yellow in Halo 10. Despite the greater number of M giants there are still a few structures only picked out in the RR Lyrae tracer as mapped in Figure \ref{fig:rrl_asp} such as the light blue and turquoise structures in the upper latitudes of Halo 10 and the dark green structure in the lower latitudes of Halo 7.
    
    In the RR Lyrae tracer we also once again see variation between the halos. Halo 12, the biggest outlier in Figure \ref{fig:rrl_cumul_present}, has several bound satellites that contain most of the mock halo's stars, while Halo 15 has mostly unbound satellites.  Halo 9 is relatively empty at high latitudes, whereas Halo 10 has two large dense structures at high latitude, and Halo 8 has a heavy distribution of stars all over. Once again stars accreted together can span 100s to 1000s of square degrees and end up on opposite ends of the halo.

\begin{figure*}
\begin{tabular}{cc}
\includegraphics[width=0.5\textwidth]{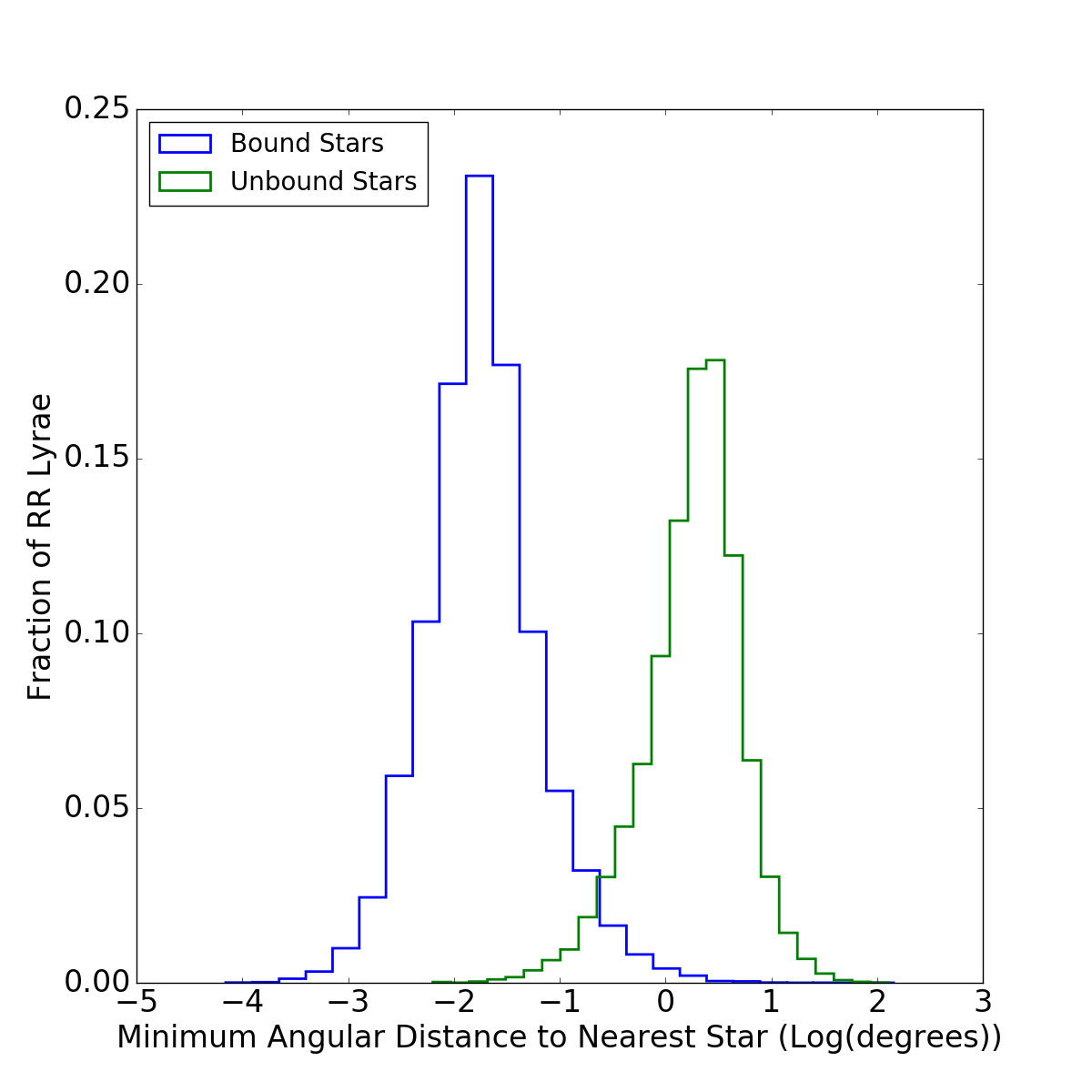} & \includegraphics[width=0.5\textwidth]{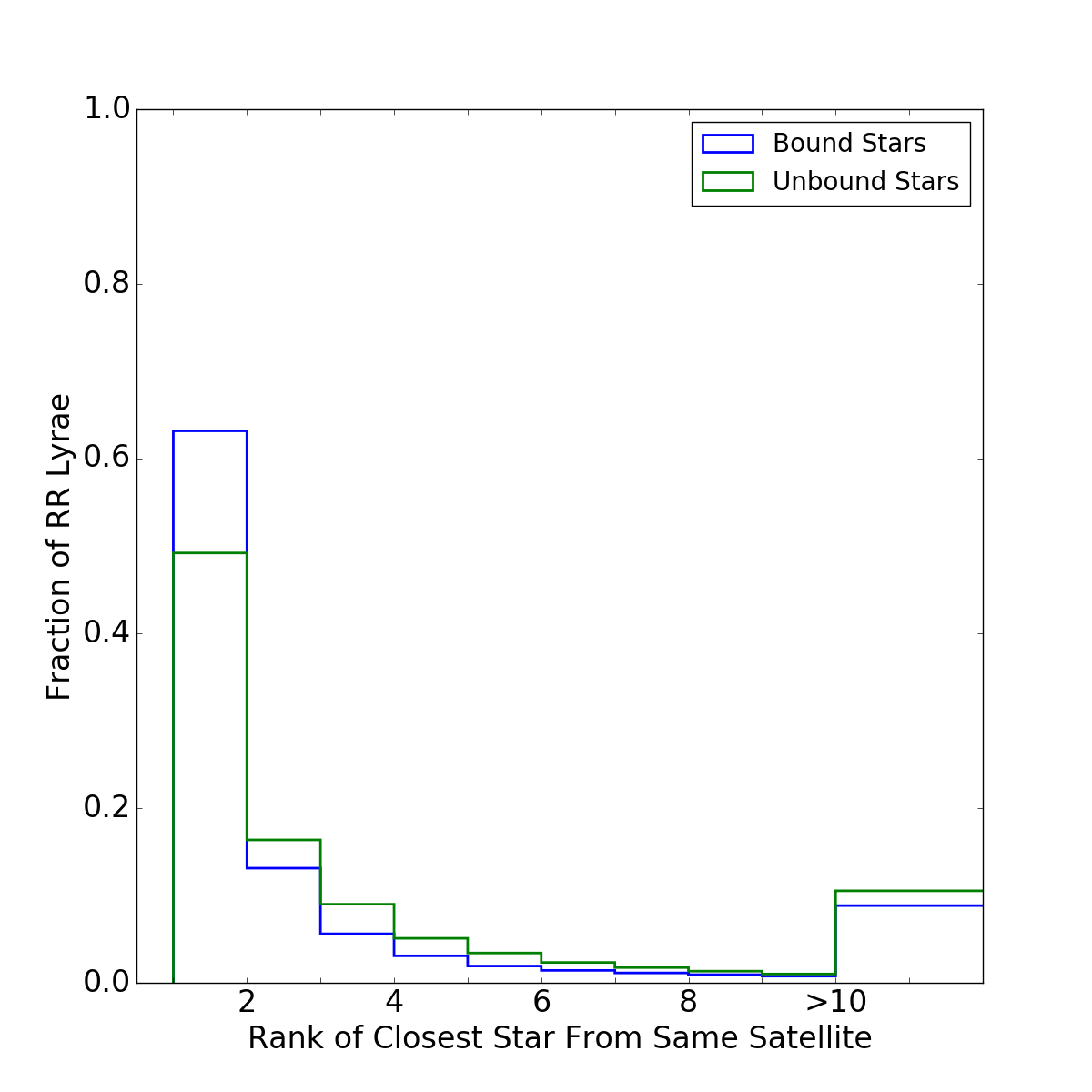} 
\end{tabular}
\caption{Left: Distribution of the minimum angular distance (in log(Degrees)) to the nearest selected RR Lyrae that was accreted as a member of the same satellite. Right: Distribution of which of the nearest selected RR Lyrae is the closest RR Lyrae that is in the same satellite.  That is 1 if the closest RR Lyrae is in the same satellite, 2 if the second closest RR Lyrae is the same satellite but the first closest is not, etc.  These distributions are for every RR Lyrae of all 11 halos between 100 to 282 kpc. Satellites that are still bound are in blue, satellites that were accreted together but have since become unbound are in green.}
\label{fig:nearestR}
\end{figure*} 

\subsection{Distribution of RR Lyrae among the accreted satellites}
\label{subsec:rrlyr-present:sats}

   As seen for M giants, there is a slight positive correlation between the number of selected RR Lyrae in a satellite and total luminosity, but with slightly more scatter at a given luminosity. This is shown in the top left panel of Figure \ref{fig:rrl_prop}, where the black line represents a linear regression of the corresponding M giant distribution in the upper left panel of Figure \ref{fig:jsat_tacc}. Unlike the M giants, since RRLe are more prominent in metal poor populations, even the lowest luminosty satellites contain a few of these rare stars, and the entire distribution is shifted towards lower luminosities for RRLe.
   
    In the remaining three panels of Figure \ref{fig:rrl_prop} we compare the properties of satellites containing 20 or more selected RR Lyrae to the those of the general population.  The typical accretion time is skewed more recently for those satellites, with a larger percentage of satellites being accreted less than 9 Gyrs ago compared to the general population. Once again this is in line with our cosmological models that suggest the outer halo was accreted more recently. The dark matter mass and luminosity of satellites containing our threshold of RR Lyrae is also skewed higher --- partially a consequence of our threshold of 20 RR Lyrae. While our results suggest that RR Lyrae in our mock surveys are more likely to have come from more massive and luminous structures, note that this bias is much less pronounced than for M giants since the number of RRLe per unit luminosity is actually decreasing as destroyed dwarfs become more massive and metal rich.

\begin{figure*}
\includegraphics[width=\textwidth,angle=270,scale=0.7]{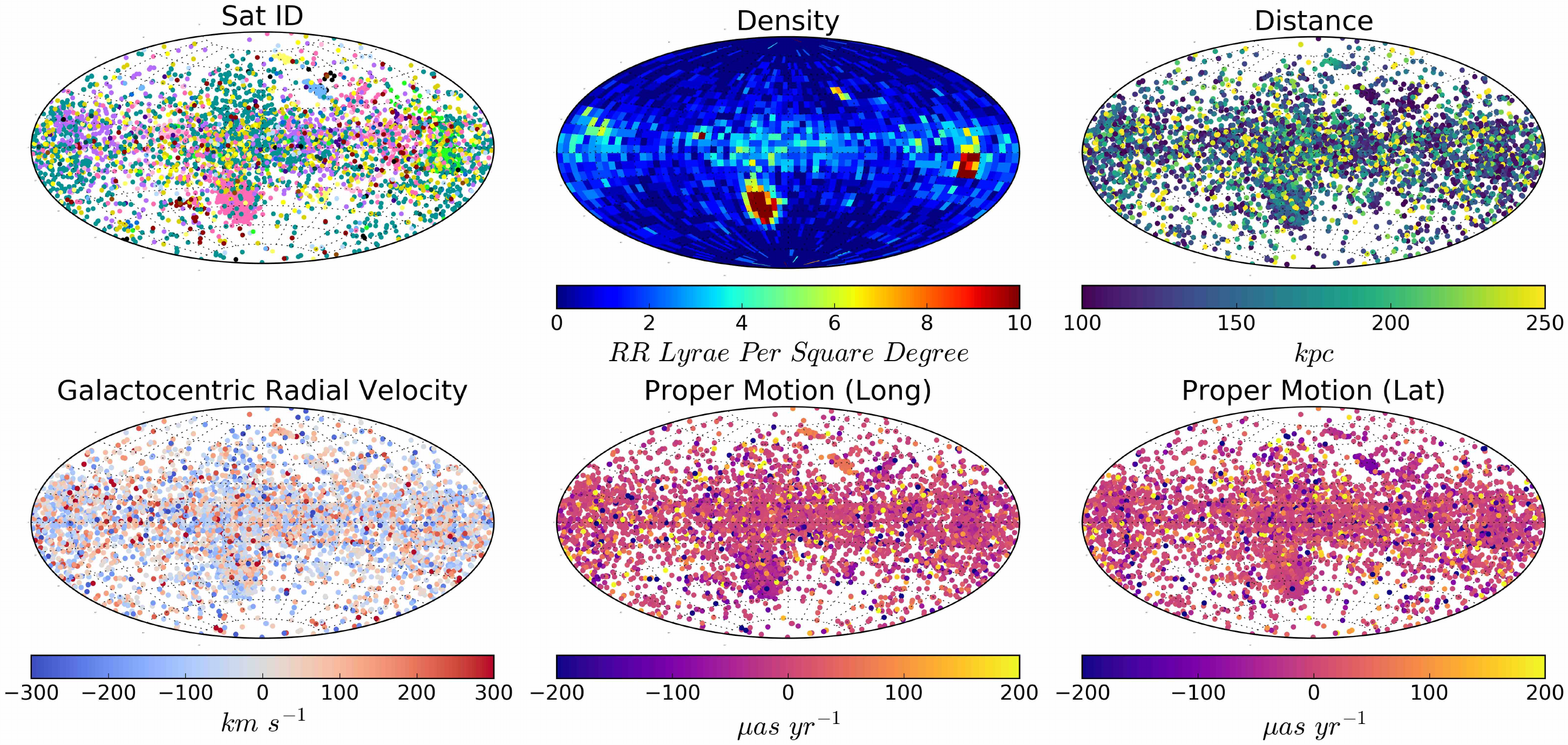}
\caption{Six all-sky views of RR Lyraes between error-convolved distances of 100 and 282 kpc for Halo 20. Each map is color coded by different properties: the satellite the star was accreted with (top left), the density, per square degree, of RR Lyrae from unbound satellites (top center), the error-convolved heliocentric distance (top right), the galactocentric radial velocity of the star (bottom left), and the galactocentric proper motion of the star (longitudinal: bottom center, latitudinal: bottom right).}
\label{fig:4x4R}
\end{figure*}

\subsection{Finding building blocks}
\label{subsec:rrl:search}
Compared to the M giants, the RR Lyrae in our mock halos seem to be a bit more well-mixed on the sky when we consider their angular distance distributions. Figure \ref{fig:nearestR} shows the distribution of minimum angular distance between two RR Lyrae from the same satellite (left), and the distribution of how many RR Lyrae away the nearest RR Lyrae from the same satellite is.  The distributions represent the selected RR Lyrae for all 11 halos combined.  The distribution of stars that are members of bound satellites are plotted in blue, and the distribution for stars that are unbound are plotted in green.  For RR Lyrae, the nearest RR Lyrae is from the same satellite roughly 60 percent of the time for bound stars, and roughly 50 percent of the time for unbound stars. This is significantly less frequent than for M giants. It is more common (over 10 percent of RR Lyrae) for the nearest RR Lyrae from the same satellite to be more than ten stars removed, for both the bound and unbound cases, than for the nearest M giant from the same satellite to be over ten stars away.  It is unsurprising then that for bound stars in particular, RR Lyrae tend to be somewhat farther from their nearest neighbor from the same satellite than M giants, with the RR Lyrae distribution peaking at around an order of magnitude larger angular distance than the M giant distribution in Figure \ref{fig:nearestM}.  For the unbound case, although both distributions peak at around a few degrees, nearly a tenth of M giants are located less than a tenth of a degree from their nearest neighbor from the same satellite, versus only a couple percent of RR Lyrae.

\begin{figure*}
\begin{tabular}{cc}
\includegraphics[width=0.5\textwidth]{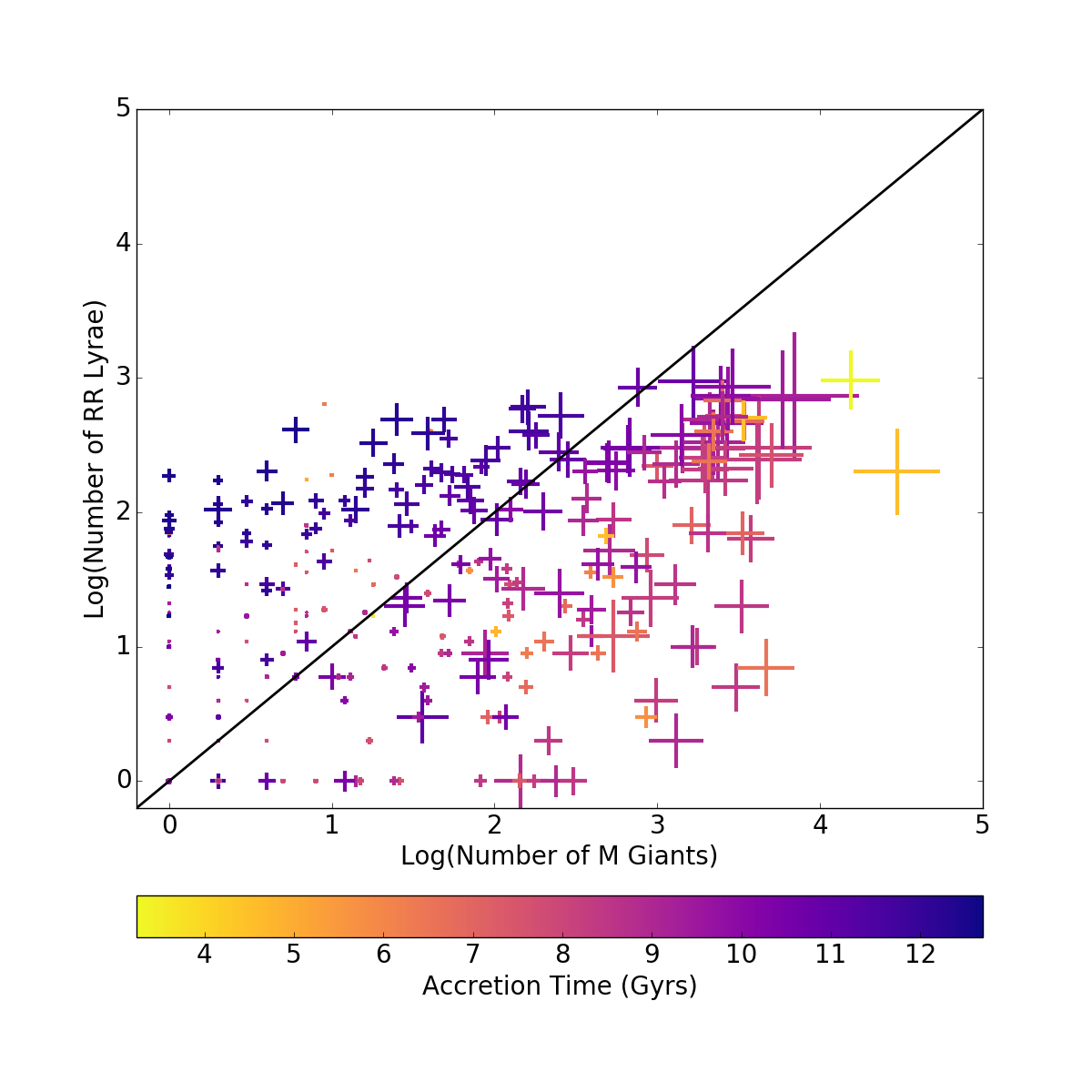} & \includegraphics[width=0.5\textwidth]{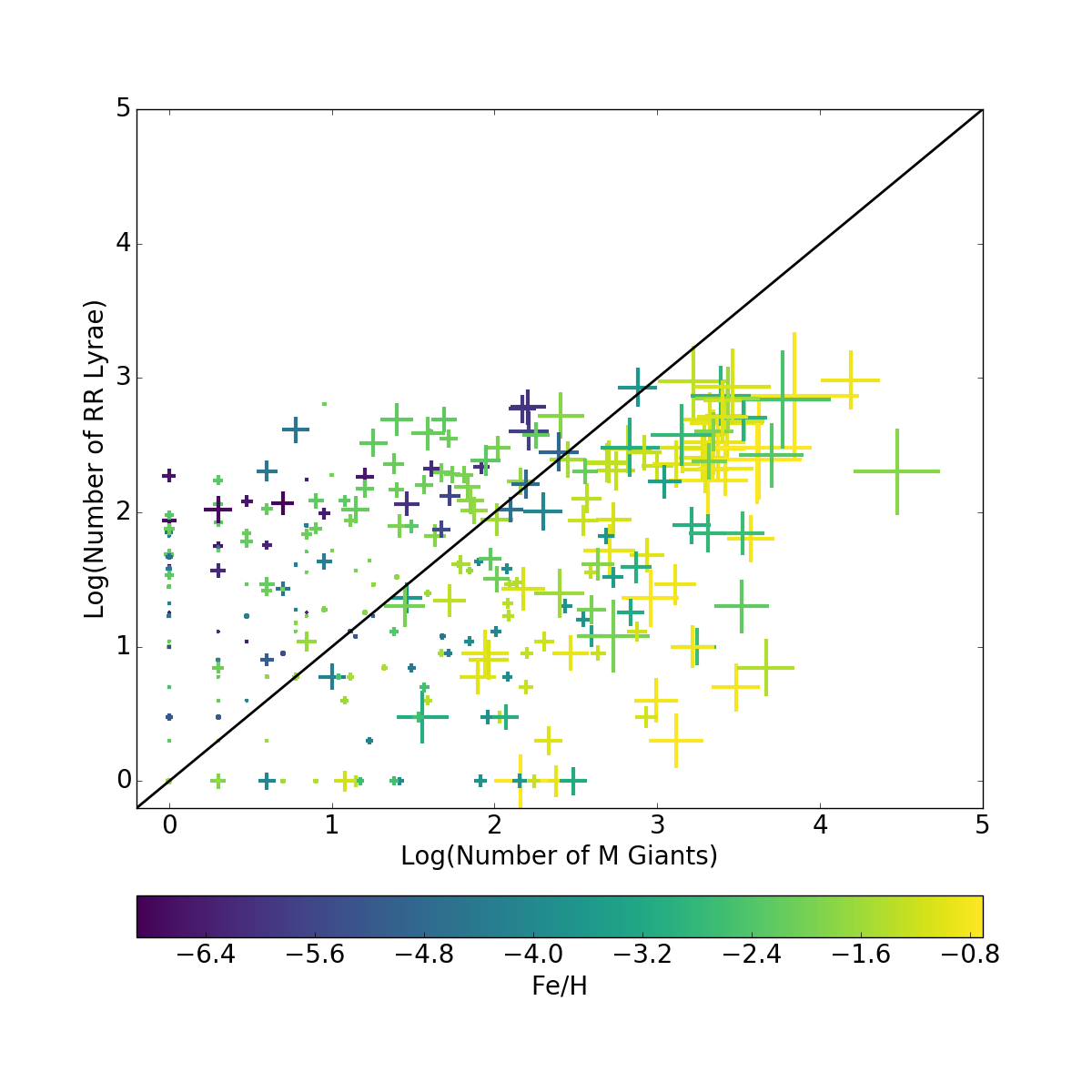} 
\end{tabular}
\caption{The distribution of the number of selected unbound M giants versus selected unbound RR Lyrae in each accreted satellite. The black diagonal line represents a one-to-one ratio. The scale of each point is a function of total satellite luminosity. The points are color coded by the how long ago the satellite was accreted in the left panel and average [Fe/H] in the right panel.}
\label{fig:rvm_scat}
\end{figure*}

    For RR Lyrae, we still see a clear distinction between bound and unbound stars in the minimum angular distance between two RR Lyrae of the same satellite and this can cautiously be used to determine if two adjacent RR Lyrae are members of a bound or unbound structure.  \citet{2015AJ....150..160B} have performed a more extensive test of the ability to detect bound structures using pairs of RR Lyrae with physical linking lengths in 2 dimensions in a series of bins in distance modulus; our finding lends support to their approach. Furthermore, although \citeauthor{2015AJ....150..160B} focused on finding bound structures (i.e. dwarf galaxies) and used an appropriately small linking length, the same technique with a larger link length could potentially also identify unbound structures. Additionally, for a survey with limited field size, our models suggest it is not only easier to locate multiple M giants together, but also that the M giants are more likely to be from the same accretion satellite, whether or not they are still bound to it.

\subsection{Untangling building blocks}

    As for the M giants, we examine which information is most useful to separate different structures from each other when viewed in RR Lyrae. Six all-sky maps of the selected RR Lyraes of Halo 20 are presented in Figure \ref{fig:4x4R}. They are color-coded as following: satellite the star was accreted with (top left), stellar density binned by square degree and only including unbound satellites (top center), error-convolved heliocentric distance (top right), galactocentric radial velocity (bottom left), and galactocentric proper motion (longitudinal: bottom center, latitudinal: bottom right). Once again we can clearly see over-densities where there are substructures, even if they are unbound. In particular the satellite colored in pink near the center of the halo and to the right, as well as the bright green satellite concentrated towards the right of the map, and the light blue satellite concentrated at higher latitude appear to be picked out nicely as over-densities. As expected, compared with the density plot in Figure \ref{fig:4x4M} there are lower densities of RR Lyrae present especially at lower latitudes.

    In Figure \ref{fig:4x4R} both longitudinal and latitudinal components of proper motion seem to be more consistently closer to zero $\mu$as/year than for the M giants in Figure \ref{fig:4x4M}. On the other hand, in particular the longitudinal component of proper motion of RR Lyrae seem to be more distinctive among different accreted satellites than in the M giant tracer, which is apparent when observing the difference in longitudinal proper motion (bottom center) between the RR Lyrae that are members of the satellite color-coded in pink in the top left plot and the other overlapping satellites. 
    
    The much smaller errors in distance for RR Lyrae lead to different structures being more uniformly colored in the distance plot (top right), suggesting in this tracer determining distances may prove more useful in differentiating between structures than for M Giants. On the other hand, there is still variation in the radial velocities of RR Lyrae from the same satellite, again because these structures are mostly shells. Overall, Figure \ref{fig:4x4R} shows that, as with M giants, over-densities can help pick out structures that can then be differentiated by proper motion components, perhaps even more effectively than in the M giant tracer and, due to smaller distance errors, by distance as well.

\begin{figure*}
\includegraphics[width=\textwidth,angle=270,scale=0.7]{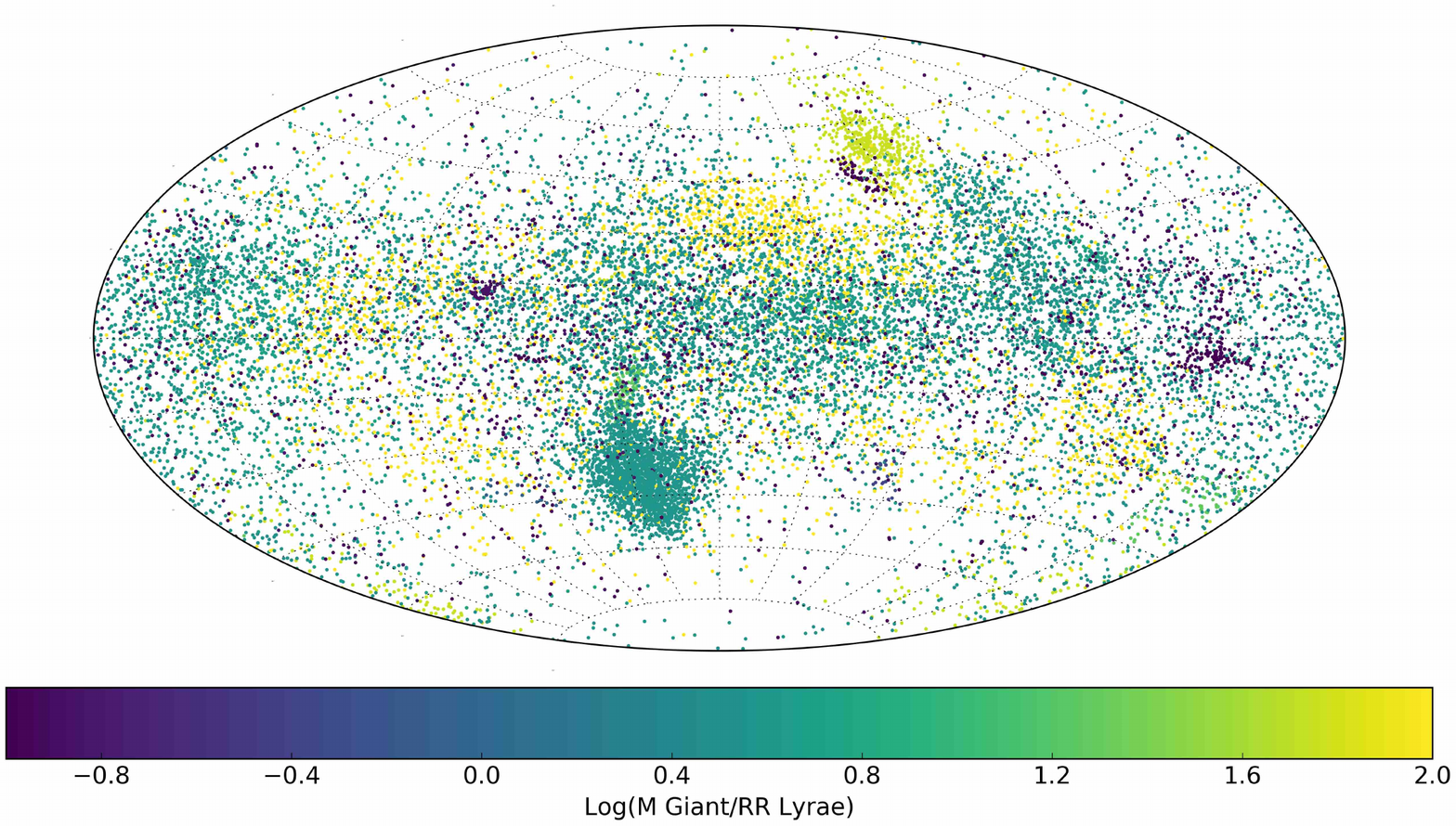}
\caption{An all sky plot of Halo 20. Each point represents either an RR Lyrae within an error convolved distance between 100 and 282 kpc or a selected M giant. Points are color-coded by the ratio of M giants to RR Lyrae in its accretion satellite. Only unbound satellites are plotted.}
\label{fig:halo20_m2r}
\end{figure*}

\section{Ratios of stellar populations}
\label{sec:ratios}
As noted by \citet{bell10}, the varying ages and metallicities of the accreted dwarf galaxies forming the distant MW halo should give rise to variations in the ratios of the abundances of different stellar populations. Our mock halos follow the star-formation model outlined in  \citet{robertson05} and \citet{font06}, which produces a correlation between accretion time and metallicity since star formation ceases at accretion, and also produces a correlation between mass (or luminosity) and metallicity via the leaky-box implementation. Furthermore, {\sc Galaxia} uses a grid of stellar populations in age and metallicity \citep{2011ApJ...730....3S} to compute the number of stellar tracers to generate. We expect that these assumptions will lead to each accreted structure in our mock halos having a different ratio of the two tracers we have studied in this work, since the relative abundance of RR Lyrae is higher at lower metallicity \citep{2015ApJ...808...50M}. We have already seen that the RR Lyrae and M giant abundances in our mock halos depend differently on luminosity, as is shown in the upper left-hand panels of Figures \ref{fig:jsat_tacc} and \ref{fig:rrl_prop}. Here we explore whether population ratios could be used to separate different accreted structures or to map out the accretion history.

First, we consider the variation in the number of M giants and RR Lyrae among the different satellites in all our mock halos. Figure \ref{fig:rvm_scat} shows that while as expected more luminous satellites (larger symbols) have more of both kind of tracer than less luminous ones, we also see that in our model the very most luminous satellites tend to have more M giants than RR Lyrae. The two panels of Figure \ref{fig:rvm_scat} illustrate how this trend arises from the star formation prescription in the model. On the left, we see the effect of truncating star formation (and hence chemical enrichment) at accretion: more recently accreted satellites have more M giants relative to RR Lyrae. On the right we see how this effect is transmitted to metallicity: the more metal-rich satellites do tend to have more M giants, but with more scatter than in accretion time. This indicates that our model should display variations in stellar populations analogous to those documented for the real halo by \citet{bell10}, and that we should be able to disentangle different accretion events (at least for these mock halos) by using ratios of stellar populations. If the model reflects the bulk properties of real chemical evolution in dwarf galaxies, we should also be able to use these ratios, combined with luminosity, to roughly date the different accretion events and reconstruct the Galactic accretion history.

Some differences between the two tracers are evident by eye when comparing the top left and center panels in Figure \ref{fig:4x4R} and Figure \ref{fig:4x4M}.  The black color-coded structure in Figure \ref{fig:4x4M} is very present and densely distributed in the M giant tracer, but looking at Figure \ref{fig:4x4R} it is barely noticeable in the RR Lyrae tracer.  There is instead a smaller over-density in the region from the small light blue structure.  In the RR Lyrae maps of Figure \ref{fig:4x4R} we also see towards the right of the map a light green structure that provides a very visible over-density in the density plot, but when looking at the M giant map, there are a few scattered green stars and no significant over-density on the right. These are multiple examples of how different structures are observable only in one or the other tracer, an argument for observing multiple tracers in order to uncover a more complete picture of the distant halo. 
    
    In the spirit of \citet{bell10}, an all-sky view of  the variations in ratio of M giants to RR Lyrae for Halo 20 is shown in Figure \ref{fig:halo20_m2r}, where each point represents a star (either an M giant or RR Lyra) color-coded by the ratio of M giants to RR Lyrae for its parent satellite. Only unbound structures are plotted. It is apparent that different structures have very different ratios of the two tracers, and that this can help distinguish overlapping structures. If some other way of determining membership was available, we could therefore use this information to reconstruct the accretion history. Based on the results of the previous sections, we tried linking overdensities in proper motion for RR Lyrae and M giants in a mixed 2000-deg${}^2$ field from one of our mock halos, and looked at whether the resulting population ratios were close to those for the parent satellites. However, this fairly simplistic search method was not successful. We attribute this on one hand to the so-called ``curse of dimensionality:'' at these large distances the structures are so sparsely sampled that in four dimensions (PMs, RV, and distance) one rapidly ends up with zero or one star per bin even if each dimension is only divided into a few bins. On the other hand, because these structures are preferentially on very radial orbits they tend to form shells rather than great-circle-like streams. This means that the although these streams are still ``cold'' in the sense that they are narrow distributions around a nearly one-dimensional path through six-dimensional phase space, the projection of that path into observed quantities is less well-confined to a few observables than for great-circle streams, where it falls primarily along the sky coordinates with small gradients in velocity and distance. For a shell the path falls mainly in the radial-velocity and distance space, but has a more complex shape in projection, and is generally double-valued in RV for a given distance. While there also exist templates for searching this space \citep{2013MNRAS.435..378S}, this points to a general need for a more sophisticated search method, so we defer this discussion to future work.

\section{Conclusions}
\label{sec:concl}

In this paper we discuss prospects for observing and untangling the distant stellar halo in two tracers: M giants and RR Lyrae. We focused on the very distant stellar halo, beyond 100 kpc, where very few stars are presently detected. Using synthetic surveys of the eleven Bullock and Johnston mock stellar halos, we found that the total number of tracers at distances beyond 100 kpc varies by about an order of magnitude from halo to halo, with most of that variation due to the number of still-bound satellites. With these removed the halo-to-halo variation is more like 0.5 dex. Excluding bound structures, our models predict roughly a few tens of thousands of M giants and a few thousand RR Lyrae beyond 100 kpc, though these absolute numbers should be taken with caution since they depend strongly on our model's assumptions about  the luminosities and stellar populations assigned to the building blocks. 

Although the different mock halos display wide variety, in general there are a few to half dozen large structures (besides the still-bound satellites) that are most prominent in a given halo and provide the best chance of identification. In quite a few cases debris from the same accreted satellite is found on opposite sides of the sky with the radial velocity signatures of shells (material piled up at apocenter), which is a result of the fairly radial orbits assigned to satellites by the model.

In studying the M giant tracers, we presumed photometric errors based on current data from the UKIDSS survey, which had the effect of scattering stars into and out of the color-color boxes used to filter out foreground dwarfs and background quasars. We found that these color cuts, combined with a very loose proper motion selection, were effective in removing self-contamination by halo dwarfs (\citet{2014AJ....147...76B} expect roughly 80 percent contamination from foreground disk dwarfs). In the future, both infrared and visible photometry should improve substantially: the WFIRST HLS will cover roughly the same sky area as UKIDSS and have significantly better IR photometric quality. While UKIDSS has sub-0.1-mag photo errors to about 18th magnitude in the J band, the WFIRST HLS will detect point sources to 5$\sigma$ down to 26.9 mag in the J band \citep{2013arXiv1305.5425S}, although it will not cover the K band. In the visible bands, LSST will see even deeper (down to 27.5 mag in the r band for 10-year coadds). Thus the technique of photometric selection should be promising for the next generation of surveys. The main challenge with so much data will be to improve the contamination by foreground disk stars, which may be possible with the improved proper motions that such surveys will also provide. 

We next considered a survey with a similar footprint size to UKIDSS, and found that a typical such footprint included stars from about $14\pm 4$ different satellites. A typical survey of this size in our mock halos is dominated by 3--5 of these satellites, each sampled by 100s--1000s of M giants. These structures could be bound or unbound, but generally appear coherent on the sky. Doubling the survey size increases the number of component satellites by less than a factor of 2, to  a median of around 18, and mainly has the effect of filling in the few most dominant structures. 

We examined which components of position and velocity might offer the best chance of disentangling different accreted structures. At these distances we found that proper motions were more diagnostic than radial velocities, which reflects the fact that most unbound structures at these distances are shells (and hence have a relatively wide range of RVs within each structure). Interestingly, the differences in proper motions between accreted structures are on the order of $\sim 50-100 \mu$as/yr, comparable to the forecasted PM uncertainties for LSST in the optical \citep{2009arXiv0912.0201L} and WFIRST in the infrared \citep{2015arXiv150303757S}. For M giants we expect distance uncertainties to be too large (about 20 percent) to be useful for disentangling tracers, but for RR Lyrae this coordinate should prove useful as well. 

Finally, we considered the prospect of untangling accretion using ratios of M giants to RR Lyrae. We found that this ratio does indeed vary substantially from one building block to another, with less luminous, older, more metal-poor satellites tending to have higher numbers of RR Lyrae relative to M giants. Given a way to confidently assign stars to structures, one could in principle use this dependence to map the accretion history of the outer halo, since the older/more metal-poor structures containing more RR Lyrae also tend to have accreted earlier. However, our preliminary attempts to separate different structures by defining one- or two-dimensional ranges in proper motion space and searching the remaining coordinates for structure were not successful, indicating that a more sophisticated strategy that takes advantage of the full multidimensionality of the available data \citep{2009ApJ...703.1061S}, and/or attempts to reduce the dimensionality to a few useful hybrid dimensions, will be needed to untangle the different accreted structures in the outer halo. Once this is done, it may be possible to then connect stars in structures on opposite sides of the sky using population ratios, which would provide valuable constraints on the Milky Way dark matter distribution at large distances.

\section*{acknowledgments}
RES is supported by an NSF Astronomy and Astrophysics Postdoctoral Fellowship under grant AST-1400989. AS contributions were also partially supported by AST-1400989. KVJ and AS contributions were enabled by NSF grant AST-1312196. JJB acknowledges support from NSF CAREER grant AST-1151462.

\bibliography{references_min} 

\begin{thebibliography}{}
\makeatletter
\relax
\def\mn@urlcharsother{\let\do\@makeother \do\$\do\&\do\#\do\^\do\_\do\%\do\~}
\def\mn@doi{\begingroup\mn@urlcharsother \@ifnextchar [ {\mn@doi@}
  {\mn@doi@[]}}
\def\mn@doi@[#1]#2{\def\@tempa{#1}\ifx\@tempa\@empty \href
  {http://dx.doi.org/#2} {doi:#2}\else \href {http://dx.doi.org/#2} {#1}\fi
  \endgroup}
\def\mn@eprint#1#2{\mn@eprint@#1:#2::\@nil}
\def\mn@eprint@arXiv#1{\href {http://arxiv.org/abs/#1} {{\tt arXiv:#1}}}
\def\mn@eprint@dblp#1{\href {http://dblp.uni-trier.de/rec/bibtex/#1.xml}
  {dblp:#1}}
\def\mn@eprint@#1:#2:#3:#4\@nil{\def\@tempa {#1}\def\@tempb {#2}\def\@tempc
  {#3}\ifx \@tempc \@empty \let \@tempc \@tempb \let \@tempb \@tempa \fi \ifx
  \@tempb \@empty \def\@tempb {arXiv}\fi \@ifundefined
  {mn@eprint@\@tempb}{\@tempb:\@tempc}{\expandafter \expandafter \csname
  mn@eprint@\@tempb\endcsname \expandafter{\@tempc}}}

\bibitem[\protect\citeauthoryear{{Abadi}, {Navarro}  \& {Steinmetz}}{{Abadi}
  et~al.}{2006}]{abadi06}
{Abadi} M.~G.,  {Navarro} J.~F.,   {Steinmetz} M.,  2006, \mnras, 365, 747

\bibitem[\protect\citeauthoryear{{Bailin}, {Bell}, {Valluri}, {Stinson},
  {Debattista}, {Couchman}  \& {Wadsley}}{{Bailin} et~al.}{2014}]{bailin14}
{Bailin} J.,  {Bell} E.~F.,  {Valluri} M.,  {Stinson} G.~S.,  {Debattista}
  V.~P.,  {Couchman} H.~M.~P.,   {Wadsley} J.,  2014, \mn@doi [\apj]
  {10.1088/0004-637X/783/2/95}, \href
  {http://adsabs.harvard.edu/abs/2014ApJ...783...95B} {783, 95}

\bibitem[\protect\citeauthoryear{{Baker} \& {Willman}}{{Baker} \&
  {Willman}}{2015}]{2015AJ....150..160B}
{Baker} M.,  {Willman} B.,  2015, \aj, 150, 160

\bibitem[\protect\citeauthoryear{{Beaton} et~al.,}{{Beaton}
  et~al.}{2016}]{2016arXiv160401788B}
{Beaton} R.~L.,  et~al., 2016, preprint, \href
  {http://adsabs.harvard.edu/abs/2016arXiv160401788B} {} (\mn@eprint {arXiv}
  {1604.01788})

\bibitem[\protect\citeauthoryear{{Bell} et~al.,}{{Bell} et~al.}{2008}]{bell08}
{Bell} E.~F.,  et~al., 2008, \apj, 680, 295

\bibitem[\protect\citeauthoryear{{Bell}, {Xue}, {Rix}, {Ruhland}  \&
  {Hogg}}{{Bell} et~al.}{2010}]{bell10}
{Bell} E.~F.,  {Xue} X.~X.,  {Rix} H.-W.,  {Ruhland} C.,   {Hogg} D.~W.,  2010,
  \aj, 140, 1850

\bibitem[\protect\citeauthoryear{{Belokurov} et~al.,}{{Belokurov}
  et~al.}{2006a}]{belokurov06}
{Belokurov} V.,  et~al., 2006a, \apj, 642, L137

\bibitem[\protect\citeauthoryear{{Belokurov} et~al.,}{{Belokurov}
  et~al.}{2006b}]{belokurov06b}
{Belokurov} V.,  et~al., 2006b, \apj, 647, L111

\bibitem[\protect\citeauthoryear{{Bertelli}, {Bressan}, {Chiosi}, {Fagotto}  \&
  {Nasi}}{{Bertelli} et~al.}{1994}]{1994A&AS..106..275B}
{Bertelli} G.,  {Bressan} A.,  {Chiosi} C.,  {Fagotto} F.,   {Nasi} E.,  1994,
  \aaps, 106

\bibitem[\protect\citeauthoryear{{Bochanski}, {Willman}, {West}, {Strader}  \&
  {Chomiuk}}{{Bochanski} et~al.}{2014a}]{2014AJ....147...76B}
{Bochanski} J.~J.,  {Willman} B.,  {West} A.~A.,  {Strader} J.,   {Chomiuk} L.,
   2014a, \aj, 147, 76

\bibitem[\protect\citeauthoryear{{Bochanski}, {Willman}, {Caldwell},
  {Sanderson}, {West}, {Strader}  \& {Brown}}{{Bochanski}
  et~al.}{2014b}]{2014ApJ...790L...5B}
{Bochanski} J.~J.,  {Willman} B.,  {Caldwell} N.,  {Sanderson} R.,  {West}
  A.~A.,  {Strader} J.,   {Brown} W.,  2014b, \apj, 790, L5

\bibitem[\protect\citeauthoryear{{Bullock} \& {Johnston}}{{Bullock} \&
  {Johnston}}{2005}]{2005ApJ...635..931B}
{Bullock} J.~S.,  {Johnston} K.~V.,  2005, \apj, 635, 931

\bibitem[\protect\citeauthoryear{{Bullock}, {Kravtsov}  \&
  {Weinberg}}{{Bullock} et~al.}{2001}]{bullock01}
{Bullock} J.~S.,  {Kravtsov} A.~V.,   {Weinberg} D.~H.,  2001, \apj, 548, 33

\bibitem[\protect\citeauthoryear{{Clementini}, {Gratton}, {Bragaglia},
  {Carretta}, {Di Fabrizio}  \& {Maio}}{{Clementini}
  et~al.}{2003}]{2003AJ....125.1309C}
{Clementini} G.,  {Gratton} R.,  {Bragaglia} A.,  {Carretta} E.,  {Di Fabrizio}
  L.,   {Maio} M.,  2003, \mn@doi [\aj] {10.1086/367773}, \href
  {http://adsabs.harvard.edu/abs/2003AJ....125.1309C} {125, 1309}

\bibitem[\protect\citeauthoryear{{Cooper} et~al.,}{{Cooper}
  et~al.}{2010}]{cooper10}
{Cooper} A.~P.,  et~al., 2010, \mnras, 406, 744

\bibitem[\protect\citeauthoryear{{De Lucia} \& {Helmi}}{{De Lucia} \&
  {Helmi}}{2008}]{delucia08}
{De Lucia} G.,  {Helmi} A.,  2008, \mnras, 391, 14

\bibitem[\protect\citeauthoryear{{Deason} et~al.,}{{Deason}
  et~al.}{2012}]{2012MNRAS.425.2840D}
{Deason} A.~J.,  et~al., 2012, \mnras, 425, 2840

\bibitem[\protect\citeauthoryear{{Drlica-Wagner} et~al.,}{{Drlica-Wagner}
  et~al.}{2015}]{drlicawagner15}
{Drlica-Wagner} A.,  et~al., 2015, \apj, 813, 109

\bibitem[\protect\citeauthoryear{{Ferguson}, {Irwin}, {Ibata}, {Lewis}  \&
  {Tanvir}}{{Ferguson} et~al.}{2002}]{ferguson02}
{Ferguson} A.~M.~N.,  {Irwin} M.~J.,  {Ibata} R.~A.,  {Lewis} G.~F.,   {Tanvir}
  N.~R.,  2002, \mn@doi [\aj] {10.1086/342019}, \href
  {http://adsabs.harvard.edu/abs/2002AJ....124.1452F} {124, 1452}

\bibitem[\protect\citeauthoryear{{Font}, {Johnston}, {Bullock}  \&
  {Robertson}}{{Font} et~al.}{2006}]{font06}
{Font} A.~S.,  {Johnston} K.~V.,  {Bullock} J.~S.,   {Robertson} B.~E.,  2006,
  \apj, 638, 585

\bibitem[\protect\citeauthoryear{{Font}, {McCarthy}, {Crain}, {Theuns},
  {Schaye}, {Wiersma}  \& {Dalla Vecchia}}{{Font} et~al.}{2011}]{font11}
{Font} A.~S.,  {McCarthy} I.~G.,  {Crain} R.~A.,  {Theuns} T.,  {Schaye} J.,
  {Wiersma} R.~P.~C.,   {Dalla Vecchia} C.,  2011, \mnras, 416, 2802

\bibitem[\protect\citeauthoryear{{Gilbert}, {Font}, {Johnston}  \&
  {Guhathakurta}}{{Gilbert} et~al.}{2009}]{gilbert09}
{Gilbert} K.~M.,  {Font} A.~S.,  {Johnston} K.~V.,   {Guhathakurta} P.,  2009,
  \apj, 701, 776

\bibitem[\protect\citeauthoryear{{Helmi}}{{Helmi}}{2004}]{2004Helmi}
{Helmi} A.,  2004, \apj, 610, L97

\bibitem[\protect\citeauthoryear{{Helmi}, {Cooper}, {White}, {Cole}, {Frenk}
  \& {Navarro}}{{Helmi} et~al.}{2011}]{helmi11}
{Helmi} A.,  {Cooper} A.~P.,  {White} S.~D.~M.,  {Cole} S.,  {Frenk} C.~S.,
  {Navarro} J.~F.,  2011, \apj, 733, L7

\bibitem[\protect\citeauthoryear{{Ibata} et~al.,}{{Ibata}
  et~al.}{2014}]{ibata14}
{Ibata} R.~A.,  et~al., 2014, \apj, 780, 128

\bibitem[\protect\citeauthoryear{{Ivezic} et~al.,}{{Ivezic}
  et~al.}{2008}]{ivezic08}
{Ivezic} Z.,  et~al., 2008, Serbian Astronomical Journal, 176, 1

\bibitem[\protect\citeauthoryear{{Ivezi{\'c}}, {Beers}  \&
  {Juri{\'c}}}{{Ivezi{\'c}} et~al.}{2012}]{ivezic12}
{Ivezi{\'c}} {\v Z}.,  {Beers} T.~C.,   {Juri{\'c}} M.,  2012, \araa, 50, 251

\bibitem[\protect\citeauthoryear{{Johnston}, {Law}  \& {Majewski}}{{Johnston}
  et~al.}{2005}]{2005Johnston}
{Johnston} K.~V.,  {Law} D.~R.,   {Majewski} S.~R.,  2005, \apj, 619, 800

\bibitem[\protect\citeauthoryear{{Johnston}, {Bullock}, {Sharma}, {Font},
  {Robertson}  \& {Leitner}}{{Johnston} et~al.}{2008}]{johnston08}
{Johnston} K.~V.,  {Bullock} J.~S.,  {Sharma} S.,  {Font} A.,  {Robertson}
  B.~E.,   {Leitner} S.~N.,  2008, \apj, 689, 936

\bibitem[\protect\citeauthoryear{{Koposov}, {Rix}  \& {Hogg}}{{Koposov}
  et~al.}{2010}]{2010Koposov}
{Koposov} S.~E.,  {Rix} H.-W.,   {Hogg} D.~W.,  2010, \apj, 712, 260

\bibitem[\protect\citeauthoryear{{K{\"u}pper}, {Balbinot}, {Bonaca},
  {Johnston}, {Hogg}, {Kroupa}  \& {Santiago}}{{K{\"u}pper}
  et~al.}{2015}]{kuepper15}
{K{\"u}pper} A.~H.~W.,  {Balbinot} E.,  {Bonaca} A.,  {Johnston} K.~V.,  {Hogg}
  D.~W.,  {Kroupa} P.,   {Santiago} B.~X.,  2015, \apj, 803, 80

\bibitem[\protect\citeauthoryear{{LSST Science Collaboration} et~al.,}{{LSST
  Science Collaboration} et~al.}{2009}]{2009arXiv0912.0201L}
{LSST Science Collaboration} et~al., 2009, preprint, \href
  {http://adsabs.harvard.edu/abs/2009arXiv0912.0201L} {} (\mn@eprint {arXiv}
  {0912.0201})

\bibitem[\protect\citeauthoryear{{Law} \& {Majewski}}{{Law} \&
  {Majewski}}{2010}]{2010Law}
{Law} D.~R.,  {Majewski} S.~R.,  2010, \apj, 714, 229

\bibitem[\protect\citeauthoryear{{Law}, {Johnston}  \& {Majewski}}{{Law}
  et~al.}{2005}]{2005Law}
{Law} D.~R.,  {Johnston} K.~V.,   {Majewski} S.~R.,  2005, \apj, 619, 807

\bibitem[\protect\citeauthoryear{{Law}, {Majewski}  \& {Johnston}}{{Law}
  et~al.}{2009}]{2009Law}
{Law} D.~R.,  {Majewski} S.~R.,   {Johnston} K.~V.,  2009, \apj, 703, L67

\bibitem[\protect\citeauthoryear{{Lux}, {Read}, {Lake}  \& {Johnston}}{{Lux}
  et~al.}{2012}]{2012MNRAS.424L..16L}
{Lux} H.,  {Read} J.~I.,  {Lake} G.,   {Johnston} K.~V.,  2012, \mnras, 424,
  L16

\bibitem[\protect\citeauthoryear{{Majewski}, {Skrutskie}, {Weinberg}  \&
  {Ostheimer}}{{Majewski} et~al.}{2003}]{majewski03}
{Majewski} S.~R.,  {Skrutskie} M.~F.,  {Weinberg} M.~D.,   {Ostheimer} J.~C.,
  2003, \apj, 599, 1082

\bibitem[\protect\citeauthoryear{{Marconi} et~al.,}{{Marconi}
  et~al.}{2015}]{2015ApJ...808...50M}
{Marconi} M.,  et~al., 2015, \mn@doi [\apj] {10.1088/0004-637X/808/1/50}, \href
  {http://adsabs.harvard.edu/abs/2015ApJ...808...50M} {808, 50}

\bibitem[\protect\citeauthoryear{{Marigo}, {Girardi}, {Bressan}, {Groenewegen},
  {Silva}  \& {Granato}}{{Marigo} et~al.}{2008}]{2008A&A...482..883M}
{Marigo} P.,  {Girardi} L.,  {Bressan} A.,  {Groenewegen} M.~A.~T.,  {Silva}
  L.,   {Granato} G.~L.,  2008, \aap, 482, 883

\bibitem[\protect\citeauthoryear{{Merrifield} \& {Kuijken}}{{Merrifield} \&
  {Kuijken}}{1998}]{1998MNRAS.297.1292M}
{Merrifield} M.~R.,  {Kuijken} K.,  1998, \mn@doi [\mnras]
  {10.1046/j.1365-8711.1998.01625.x}, \href
  {http://adsabs.harvard.edu/abs/1998MNRAS.297.1292M} {297, 1292}

\bibitem[\protect\citeauthoryear{{Newberg} et~al.,}{{Newberg}
  et~al.}{2002}]{newberg02}
{Newberg} H.~J.,  et~al., 2002, \apj, 569, 245

\bibitem[\protect\citeauthoryear{{Newberg}, {Willett}, {Yanny}  \&
  {Xu}}{{Newberg} et~al.}{2010}]{2010Newberg}
{Newberg} H.~J.,  {Willett} B.~A.,  {Yanny} B.,   {Xu} Y.,  2010, \apj, 711, 32

\bibitem[\protect\citeauthoryear{{Pearson}, {K{\"u}pper}, {Johnston}  \&
  {Price-Whelan}}{{Pearson} et~al.}{2015}]{pearson15}
{Pearson} S.,  {K{\"u}pper} A.~H.~W.,  {Johnston} K.~V.,   {Price-Whelan}
  A.~M.,  2015, \apj, 799, 28

\bibitem[\protect\citeauthoryear{{Pillepich}, {Madau}  \& {Mayer}}{{Pillepich}
  et~al.}{2015}]{pillepich15}
{Pillepich} A.,  {Madau} P.,   {Mayer} L.,  2015, \apj, 799, 184

\bibitem[\protect\citeauthoryear{{Robertson}, {Bullock}, {Font}, {Johnston}  \&
  {Hernquist}}{{Robertson} et~al.}{2005}]{robertson05}
{Robertson} B.,  {Bullock} J.~S.,  {Font} A.~S.,  {Johnston} K.~V.,
  {Hernquist} L.,  2005, \apj, 632, 872

\bibitem[\protect\citeauthoryear{{Sanderson} \& {Helmi}}{{Sanderson} \&
  {Helmi}}{2013}]{2013MNRAS.435..378S}
{Sanderson} R.~E.,  {Helmi} A.,  2013, \mnras, 435, 378

\bibitem[\protect\citeauthoryear{{Schlaufman} et~al.,}{{Schlaufman}
  et~al.}{2009}]{schlaufman09}
{Schlaufman} K.~C.,  et~al., 2009, \apj, 703, 2177

\bibitem[\protect\citeauthoryear{{Sesar} et~al.,}{{Sesar}
  et~al.}{2010}]{2010ApJ...708..717S}
{Sesar} B.,  et~al., 2010, \mn@doi [\apj] {10.1088/0004-637X/708/1/717}, \href
  {http://adsabs.harvard.edu/abs/2010ApJ...708..717S} {708, 717}

\bibitem[\protect\citeauthoryear{{Sesar} et~al.,}{{Sesar}
  et~al.}{2013}]{2013ApJ...776...26S}
{Sesar} B.,  et~al., 2013, \mn@doi [\apj] {10.1088/0004-637X/776/1/26}, \href
  {http://adsabs.harvard.edu/abs/2013ApJ...776...26S} {776, 26}

\bibitem[\protect\citeauthoryear{{Sesar} et~al.,}{{Sesar}
  et~al.}{2014}]{2014ApJ...793..135S}
{Sesar} B.,  et~al., 2014, \mn@doi [\apj] {10.1088/0004-637X/793/2/135}, \href
  {http://adsabs.harvard.edu/abs/2014ApJ...793..135S} {793, 135}

\bibitem[\protect\citeauthoryear{{Sharma} \& {Johnston}}{{Sharma} \&
  {Johnston}}{2009}]{2009ApJ...703.1061S}
{Sharma} S.,  {Johnston} K.~V.,  2009, \apj, 703, 1061

\bibitem[\protect\citeauthoryear{{Sharma}, {Johnston}, {Majewski}, {Bullock}
  \& {Mu{\~n}oz}}{{Sharma} et~al.}{2011a}]{2011ApJ...728..106S}
{Sharma} S.,  {Johnston} K.~V.,  {Majewski} S.~R.,  {Bullock} J.,   {Mu{\~n}oz}
  R.~R.,  2011a, \apj, 728, 106

\bibitem[\protect\citeauthoryear{{Sharma}, {Bland-Hawthorn}, {Johnston}  \&
  {Binney}}{{Sharma} et~al.}{2011b}]{2011ApJ...730....3S}
{Sharma} S.,  {Bland-Hawthorn} J.,  {Johnston} K.~V.,   {Binney} J.,  2011b,
  \apj, 730, 3

\bibitem[\protect\citeauthoryear{{Spergel} et~al.,}{{Spergel}
  et~al.}{2013}]{2013arXiv1305.5425S}
{Spergel} D.,  et~al., 2013, preprint, \href
  {http://adsabs.harvard.edu/abs/2013arXiv1305.5425S} {} (\mn@eprint {arXiv}
  {1305.5425})

\bibitem[\protect\citeauthoryear{{Spergel} et~al.,}{{Spergel}
  et~al.}{2015}]{2015arXiv150303757S}
{Spergel} D.,  et~al., 2015, preprint, \href
  {http://adsabs.harvard.edu/abs/2015arXiv150303757S} {} (\mn@eprint {arXiv}
  {1503.03757})

\bibitem[\protect\citeauthoryear{{Tissera}, {Scannapieco}, {Beers}  \&
  {Carollo}}{{Tissera} et~al.}{2013}]{tissera13}
{Tissera} P.~B.,  {Scannapieco} C.,  {Beers} T.~C.,   {Carollo} D.,  2013,
  \mnras, 432, 3391

\bibitem[\protect\citeauthoryear{{VanderPlas} \& {Ivezi\v c}}{{VanderPlas} \&
  {Ivezi\v c}}{2015}]{2015ApJ...812...18V}
{VanderPlas} J.~T.,  {Ivezi\v c} {\v Z}.,  2015, \apj, 812, 18

\bibitem[\protect\citeauthoryear{{Vera-Ciro} \& {Helmi}}{{Vera-Ciro} \&
  {Helmi}}{2013}]{2013Vera-Ciro}
{Vera-Ciro} C.,  {Helmi} A.,  2013, \apj, 773, L4

\bibitem[\protect\citeauthoryear{{Wetzel}, {Hopkins}, {Kim},
  {Faucher-Gigu{\`e}re}, {Kere{\v s}}  \& {Quataert}}{{Wetzel}
  et~al.}{2016}]{2016ApJ...827L..23W}
{Wetzel} A.~R.,  {Hopkins} P.~F.,  {Kim} J.-h.,  {Faucher-Gigu{\`e}re} C.-A.,
  {Kere{\v s}} D.,   {Quataert} E.,  2016, \mn@doi [\apj]
  {10.3847/2041-8205/827/2/L23}, \href
  {http://adsabs.harvard.edu/abs/2016ApJ...827L..23W} {827, L23}

\bibitem[\protect\citeauthoryear{{Willett}, {Newberg}, {Zhang}, {Yanny}  \&
  {Beers}}{{Willett} et~al.}{2009}]{2009Willett}
{Willett} B.~A.,  {Newberg} H.~J.,  {Zhang} H.,  {Yanny} B.,   {Beers} T.~C.,
  2009, \apj, 697, 207

\bibitem[\protect\citeauthoryear{{Willman} et~al.,}{{Willman}
  et~al.}{2005}]{willman05}
{Willman} B.,  et~al., 2005, \apj, 626, L85

\bibitem[\protect\citeauthoryear{{Xue} et~al.,}{{Xue} et~al.}{2011}]{xue11}
{Xue} X.-X.,  et~al., 2011, \apj, 738, 79

\bibitem[\protect\citeauthoryear{{Zolotov}, {Willman}, {Brooks}, {Governato},
  {Brook}, {Hogg}, {Quinn}  \& {Stinson}}{{Zolotov} et~al.}{2009}]{zolotov09}
{Zolotov} A.,  {Willman} B.,  {Brooks} A.~M.,  {Governato} F.,  {Brook} C.~B.,
  {Hogg} D.~W.,  {Quinn} T.,   {Stinson} G.,  2009, \apj, 702, 1058

\makeatother
\end{thebibliography}

\bsp
\label{lastpage}

\end{document}